\documentclass[12pt,a4paper]{article}
\pdfoutput=1

\usepackage{jheppub}
\usepackage{color}
\usepackage{tikz}
\usepackage{bm}
\usepackage{amsmath}
\usepackage{amssymb}
\usepackage{amsthm}
\usepackage{ccaption}
\usepackage{subcaption}
\usepackage{graphicx}
\usepackage[super]{nth}
\usepackage{microtype}
\usepackage{cleveref}
\usepackage{listings}
\usepackage{dsfont}

\graphicspath{{Figures/}}

  \captionnamefont{\bfseries}
  \captiontitlefont{\small\sffamily}
  \captiondelim{: }
  \hangcaption

\DeclareMathOperator{\csch}{csch}

\newcommand{\be}{\begin{equation}}
\newcommand{\ee}{\end{equation}}
\newcommand{\beq}{\begin{eqnarray}}
\newcommand{\eeq}{\end{eqnarray}}

\author[a]{Donald Marolf\,}
\author[b]{\!, Henry Maxfield\,}
\author[b]{\!, Alex Peach\,}
\author[b]{\!, Simon Ross}
\affiliation[\ a]{Department of Physics,\\
                     University of California, Santa Barbara, Santa Barbara, CA 93106, USA.}
\affiliation[\,b]{Centre for Particle Theory \& Department of Mathematical Sciences,\\
                     Durham University, South Road, Durham DH1 3LE, UK.}

\emailAdd{marolf@physics.ucsb.edu}
\emailAdd{h.d.maxfield@durham.ac.uk}
\emailAdd{s.f.ross@durham.ac.uk}
\emailAdd{a.m.peach@durham.ac.uk}

\abstract{
We analyze the 1+1 CFT states dual to hot (time-symmetric) 2+1 multiboundary AdS wormholes. These are black hole geometries with high local temperature,  $n \ge 1$ asymptotically-AdS$_3$ regions, and arbitrary internal topology. The dual state at $t=0$ is defined on $n$ circles.  We show these to be well-described by sewing together tensor networks corresponding to thermofield double states.  As a result, the entanglement is spatially localized and bipartite: away from particular boundary points (``vertices'') any small connected region $A$ of the boundary CFT  is entangled only with another small connected region $B$, where $B$ may lie on a different circle or may be a different part of the same circle.  We focus on the pair-of-pants case, from which more general cases may be constructed.  We also discuss finite-temperature corrections, where we note that the states involve a code subspace in each circle.}


\begin{document}
\title{Hot multiboundary wormholes from bipartite entanglement}


\maketitle

\flushbottom
\renewcommand{\thefootnote}{\arabic{footnote}}

\section{Introduction}

The thermofield double (TFD) state
\begin{equation}
\label{eq:TFD}
|TFD \rangle = \sum_E e^{-E/2T} |E \rangle |E \rangle
\end{equation}
on two copies of a quantum field theory serves as the poster child for many ideas
\cite{Maldacena:2001kr,VanRaamsdonk:2010pw,Czech:2012be,Maldacena:2013xja,Hubeny:2007xt,Lewkowycz:2013nqa,Ryu:2006bv} relating the emergence of bulk geometry to entanglement in some dual theory.  As explained in \cite{Maldacena:2001kr}, although a single copy of a CFT can be naturally dual to bulk quantum gravity with a single asymptotically AdS boundary, the particular entanglements between the two copies described by \eqref{eq:TFD} allow it to be dual to a two-sided eternal black hole in which two distinct asymptotic regions are connected by an Einstein-Rosen bridge\footnote{Though there may be interesting subtleties; see e.g. \cite{Marolf:2012xe}.}.  The state also typifies relations between the area of codimension-2 surfaces and CFT entanglement encapsulated in the Ryu-Takayangi conjecture \cite{Ryu:2006bv} and the covariant generalization by Hubeny, Rangamani, and Takayanagi (HRT) \cite{Hubeny:2007xt}.   Here and below we work in the regime where the bulk planck scale $\ell_p$ is small in comparison with the bulk AdS scale $\ell_{AdS}$ (which we generally set to 1), or equivalently where $N \gg 1$ in the CFT (i.e., large central charge $c$ for a $1+1$ CFT).

In discussing $|TFD\rangle$, it is natural to focus on the bipartite entanglement between the associated two copies of the CFT.  This entanglement has a special structure:  as shown in \cite{Morrison:2012iz}, the entanglement is both local and bipartite in the sense that, when studying regions of the CFT of size greater than the thermal scale, a given region can be said to be entangled {\it only} with the corresponding spatial region in the second CFT.  In particular, when we consider regions $A$, $B$ (in the same or opposite CFTs) separated by more than this scale, the mutual information
\begin{equation}
I(A : B) = S(A) + S(B) - S(A B),
\end{equation}
vanishes at leading order in large $N$. This result can easily be understood from a CFT path integral point of view. In general, the thermofield double state is calculated by a CFT path integral over a cylinder, linking the two copies of the spatial section the state \eqref{eq:TFD} is defined on. In the high temperature limit, this cylinder becomes short compared to its circumference, so when we consider regions larger than the length of the cylinder, the resulting state naturally only entangles regions on one boundary with the corresponding region on the other boundary.

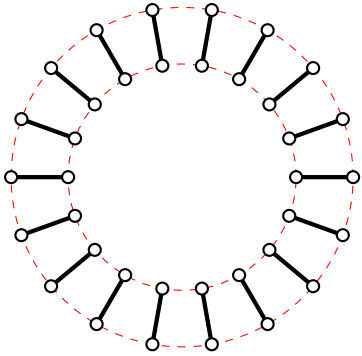
\begin{figure}
\centering
	\begin{tikzpicture}[scale=1.5]
		\draw [red, dashed] (0,0) circle [radius=1.5];
		\draw [red, dashed] (0,0) circle [radius=1];
		\foreach \x in {1,...,18}
			{\draw [ultra thick] ({cos(20*\x)},{sin(20*\x)} ) -- ({1.5*cos(20*\x)},{1.5*sin(20*\x)} );
			\draw [thick, fill=white] ({cos(20*\x)},{sin(20*\x)} ) circle (1.5pt);
			\draw[thick, fill=white] ({1.5*cos(20*\x)},{1.5*sin(20*\x)} ) circle (1.5pt);}
	\end{tikzpicture}
\caption{A simple tensor network displaying the localized purely-bipartite entanglement characteristic of holographic $|TFD\rangle$ states at large $N$ on scales longer than the thermal scale. Each node represents a region in the CFT of scale longer than the thermal scale.  We focus mainly on CFT states on $S^1 \times {\mathbb R}$ where one takes a high-temperature limit in order to fit many such long-distance regions onto the circle, though one may equally-well consider the planar case. The solid links are the entangling tensors implied by \eqref{eq:TFD}.  The dashed lines guide the eye by linking neighbouring regions in each of the two CFTs.}
\label{fig:TFDnet}
\end{figure}

It will be useful below to visualize this result in the language of tensor networks; see e.g. \cite{2009arXiv0912.1651V}.  The rather trivial nature of the above entanglement then translates into a similarly-trivial coarse-grained tensor network description of $|TFD \rangle$ as shown in figure \ref{fig:TFDnet}.

While the thermofield double state is a useful simple example, it is important to find further examples where we can understand the relation of bulk geometry to CFT entanglement structures. We are also interested in exploring the role played by multi-party entanglement in connections between 3 or more subsystems and what form it takes in the associated CFT states, see e.g. \cite{Susskind:2014ira,Balasubramanian:2014hda,Susskind:2014yaa}.

The vast literature on holographic entanglement has focused primarily on bipartite relations between a given subsystem in the CFT and its complement, so that relatively little is known about multiparty issues.  One general result is the monogamy of holographic entanglement established in \cite{Hayden:2011ag}.  But a more detailed investigation of multipartite entanglement was recently initiated in \cite{Balasubramanian:2014hda} using a class of 2+1-dimensional black hole spacetimes \cite{Brill:1995jv,Aminneborg:1997pz,Brill:1998pr,Aminneborg:1998si} describing a collapsing wormhole that connects $n$ regions each asymptotic to (global) AdS${}_3$.  When the corresponding Euclidean geometries define the dominant saddle of a natural path integral, such geometries are dual to entangled states on $n$ copies of a 1+1 dimensional CFT on $S^1 \times {\mathbb R}$ defined by a path integral on a Riemann surface $\Sigma$ with $n$ circular boundaries \cite{Maldacena:2001kr,Krasnov:2000zq,Krasnov:2003ye,Skenderis:2009ju}.  The corresponding entanglement was found to display a rich dependence on the moduli, including regimes of purely bipartite entanglement, and others of strong multipartite entanglement.  Interestingly, the strongly multipartite regions identified in \cite{Balasubramanian:2014hda} corresponded to bulk black holes with temperature less than the AdS scale\footnote{It remains an open question whether such phases ever dominate the path integrals described above.  But even if not, one presumes them to be dual to some other class of CFT states whose entanglement must be correspondingly multipartite.}.  The recent work \cite{Bao:2015bfa} describes an infinite family of generalizations of results from both \cite{Hayden:2011ag} and \cite{Balasubramanian:2014hda}.

\begin{figure}
\centering
\begin{subfigure}[t]{0.4\textwidth}
\centering
\includegraphics[scale=0.8]{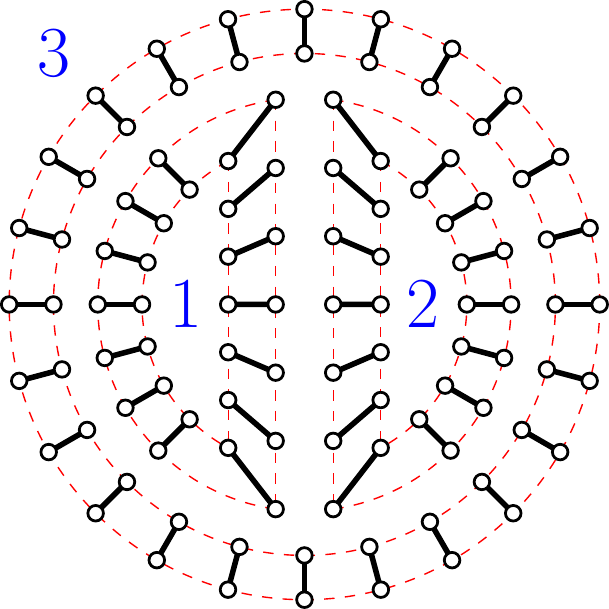}
\end{subfigure}
\hspace{.08\textwidth}
\begin{subfigure}[t]{0.4\textwidth}
\centering
\includegraphics[scale=0.8]{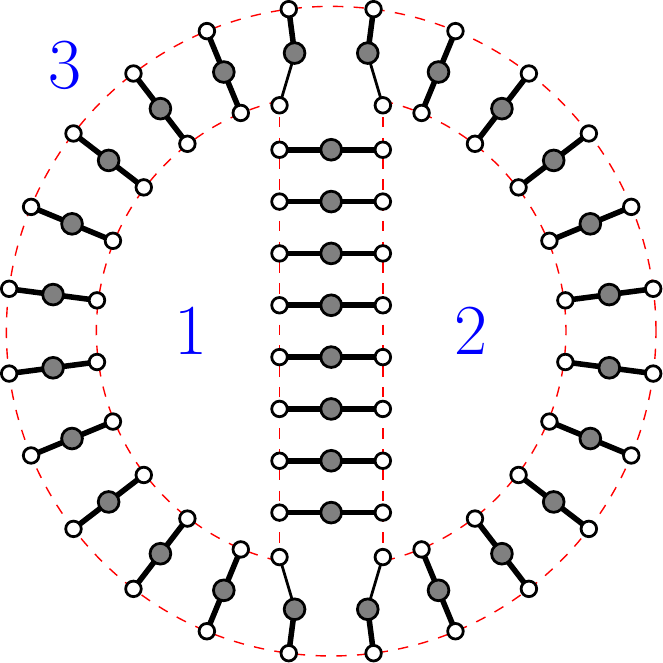}
\end{subfigure}\\
\begin{subfigure}[t]{0.43\textwidth}
\centering
\includegraphics[scale=0.8]{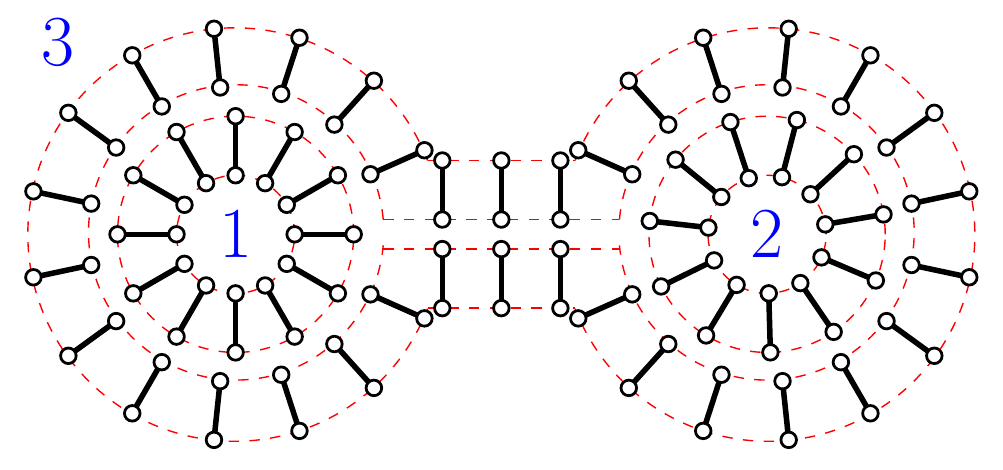}
\end{subfigure}
\hspace{.02\textwidth}
\begin{subfigure}[t]{0.5\textwidth}
\centering
\includegraphics[scale=0.8]{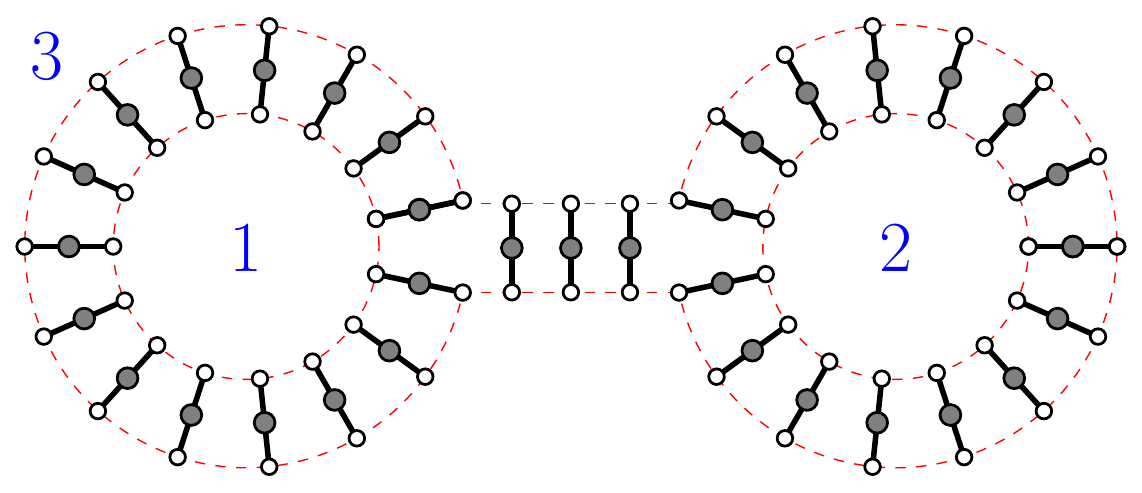}
\end{subfigure}
\caption{Two topologically-distinct ways in which three copies of the tensor network in figure \ref{fig:TFDnet} can be sewn together (left figures) into a single tensor network (right figures) defining a state on 3 copies of the system.  The dashed lines (red in color version) internal to the left diagrams guide the eye toward recognising the 3 constituent copies of the network in figure \ref{fig:TFDnet}.  Links that meet across adjoining pairs of dashed lines are contracted, establishing entanglement between the remaining boundaries (marked 1, 2, and 3).  In the bottom-left figure, two parts of the outermost tensor network are contracted with each other, resulting in two well-separated regions of boundary 1 becoming entangled with each other as shown in the bottom-right figure.  As discussed below, all 3-boundary time-symmetric vacuum wormholes with pair-of-pants topology (orientable with no handles) and large horizons correspond at the moment of time-symmetry to one of the cases shown, or to the degenerate case that interpolates between them, when described in the ``round'' conformal frame in which the energy density is taken to be constant along each of the 3 boundaries.  Although we show only a simplified cartoon of the full tensor network, we argue below that sewing the actual $|TFD\rangle$ tensor networks together in this way describes the corresponding CFT states with exponential accuracy away from the two `vertices' in each diagram where 3 $|TFD\rangle$'s meet.
}
\label{fig:sew}
\end{figure}

We focus below on the opposite limit in which all bulk black holes have high temperature. The length of their horizons is then very large with respect to the AdS scale.  We will show that these geometries are dual to states constructed by sewing together copies of $|TFD\rangle$, as shown in figure  \ref{fig:sew}. The entanglement is thus both local and bipartite away from small regions containing certain ``vertices" where the sewing involves three or more copies of $|TFD\rangle$. From the CFT path integral point of view, this arises because the boundary circles are large compared to the distance between them; in a conformal frame where the boundaries are finite size, there are thin strips joining them, corresponding to the short tensor networks in figure \ref{fig:sew}. In section \ref{sec:CFT} we will justify this picture more quantitatively by showing that local pieces of the surface $\Sigma$ are described by regions of BTZ up to exponential corrections.  As a result, as in \cite{Pastawski:2015qua} tripartite entanglement appears to localise in isolated AdS-scale regions of the bulk.  Away from these vertices, the construction of the state involves only the sewing together $|TFD\rangle$'s of inverse temperature $\beta_1$ and $\beta_2$, giving a local version of the $|TFD\rangle$ of inverse temperature $\beta_1 + \beta_2$.  Since we focus on 1+1 CFTs, we henceforth refer to the limit of large central charge $c$ rather than large $N$.

Note that nothing prevents sewing operations that link together disjoint regions in the same CFT as shown in figure \ref{fig:sew} (bottom).  As we will see, this also provides an interesting picture in our limit of CFT states dual to single-boundary black holes with internal topology.  The reader should thus be aware that, while we use term ``multiboundary'' below, this explicitly includes the very interesting case $n=1$ as well as $n\ge 2$.

One may expect each local piece of $|TFD\rangle$ in figure \ref{fig:sew} (right) to correspond to a bulk region whose geometry near $t=0$ is well-approximated by a corresponding piece of BTZ.  We show in section  \ref{holo} that this is indeed the case, and thus that bulk  Ryu-Takayanagi (or, more precisely, HRT) calculations are consistent with the entanglements shown.

We begin by reviewing aspects of general multiboundary wormholes and their relation to CFT states in section \ref{review}.  Section \ref{sec:LargeL} then studies the  high temperature (equivalently, large horizon length $L$) limit of the geometry of $\Sigma$.  We show that the region between the horizons in $\Sigma$ becomes unimportant in this limit.  This allows us to argue in section \ref{sec:CFT} that the CFT path integral produces the structure described by figure \ref{fig:sew}. Section \ref{holo} then describes how this same result is seen in the bulk HRT calculation and argues that the desired bulk wormhole does indeed dominate the corresponding bulk path integral. Section \ref{finite} briefly addresses finite temperature corrections and we conclude in section \ref{disc} with discussions of the general $n$-boundary case, internal topology, higher dimensions, and future directions.  In particular we comment explicitly on examples with $n=1$.

\section{Path integrals, states, and bulk geometries}
\label{review}

We now commence our review.  As is well known, the thermofield double state of inverse temperature $\beta$ is computed by the CFT path integral on the cylinder of circumference $2\pi$ and height $\beta/2$.  Here one regards each of the two circular boundaries as the (say) $t=0$ slice of a corresponding CFT.  The path integral between field configurations $\phi_1, \phi_2$ on the two boundaries then gives the wavefunction $\Psi(\phi_1, \phi_2)$ of the joint state of the two CFTs.\footnote{In this discussion we assume for simplicity that the CFTs admit (anti-unitary) time-reversal symmetries $T$ which can be used to map bra-vectors to ket-vectors and vice versa, and which can therefore be used to construct \eqref{eq:TFD} from the thermal operator $e^{-\frac{1}{2}\beta H}$.} At sufficiently high temperatures, the corresponding bulk path integral is dominated by a saddle point associated with the Euclidean BTZ black hole.  In this case we may say that, to good approximation, the corresponding Lorentz-signature bulk black hole is dual to $|TFD \rangle$.

\begin{figure}
\centering
\includegraphics[keepaspectratio,width=0.5\linewidth]{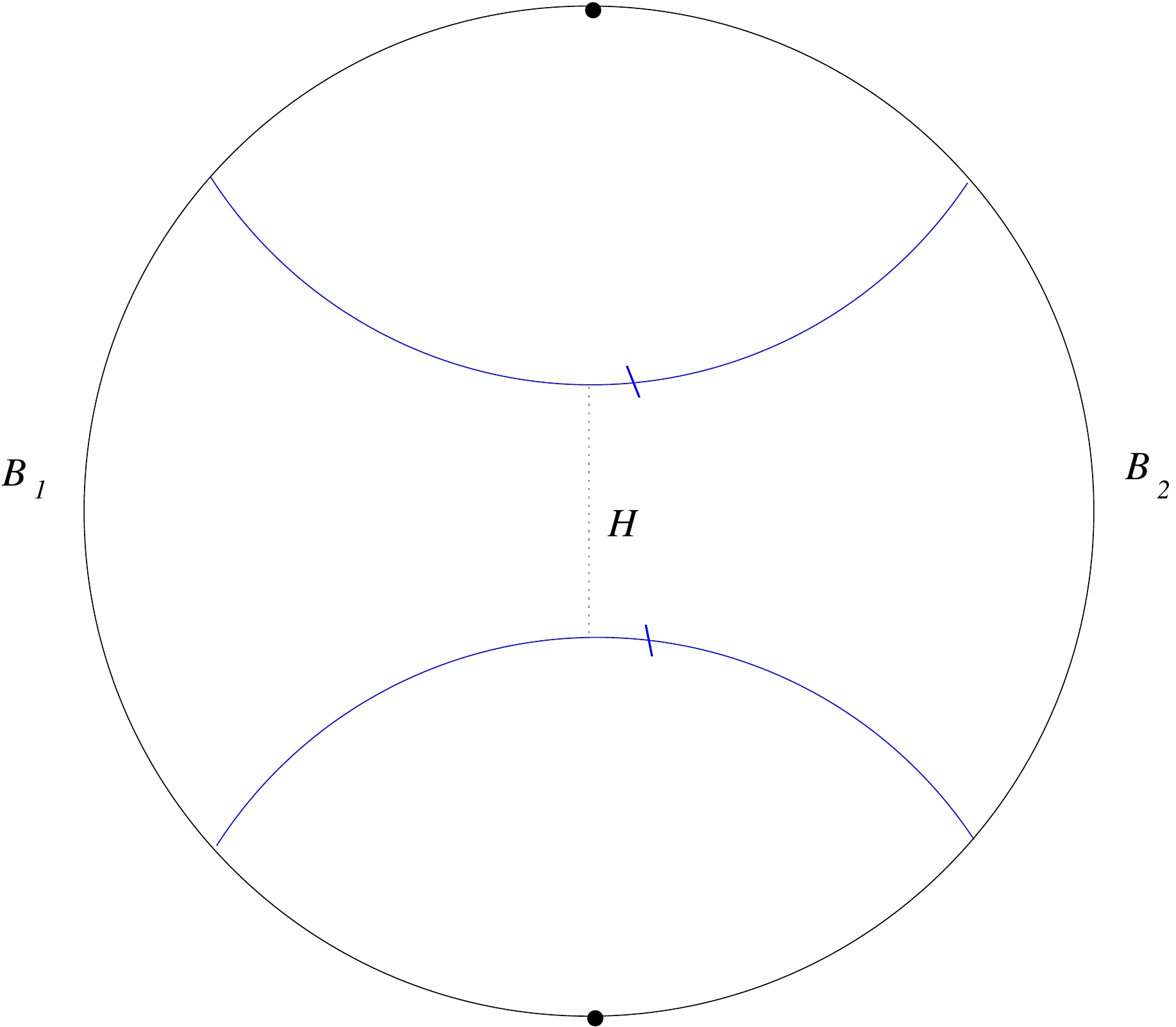}
\caption{The quotient of the hyperbolic plane $H^2$ by $\Gamma$. The pair of labeled geodesics are identified by $g$, so the region between them forms a fundamental domain for the quotient. The minimal closed geodesic $H$ is the horizon for the resulting BTZ geometry.}
\label{cyl}
\end{figure}

The cylinder of circumference $2\pi$ and height $\beta/2$ plays two important roles in the BTZ geometry.  First, it is conformally equivalent to (half of) the boundary of Euclidean BTZ.  This is what allows Euclidean BTZ to be a saddle for the desired bulk path integral.  But this same cylinder is also conformal to the BTZ geometry at $t=0$, which may be equally-well considered as a slice of either the Euclidean or the Lorentzian black hole.  This may be seen by recalling \cite{Banados:1992gq} that Euclidean BTZ can be constructed as a quotient of global Euclidean AdS${}_3$ (i.e., of the hyperbolic three-space $H^3$) by an isometry.  The simplest statement requires two steps.  One first writes Euclidean AdS${}_3$ in terms of its slicing by hyperbolic planes $H^2$ (equivalently, by copies of Euclidean AdS$_2$) as
\begin{equation}
\label{eq:hypslice}
\frac{ds^2}{\ell_{AdS}^2} =  dt_E^2 + \cosh^2 t_E d \Sigma^2,
\end{equation}
where   $d \Sigma^2$ is the metric on the unit-radius $H^2$.  One then quotients each $H^2$ slice by a discrete group $\Gamma = \{g^n : n \in {\mathbb Z} \}$ generated by some hyperbolic element $g$ of its $SL(2,{\mathbb R})$ group.\footnote{In other words, thinking of $SL(2,{\mathbb R})$ as the Lorentz group $SO(2,1)$ of 2+1 Minkowski space this $g$ must be a boost preserving some spacelike direction.}  The action of $g$ and its fundamental domain in $H^2$ are indicated in figure \ref{cyl}.  Since the different $H^2$ slices in \eqref{eq:hypslice} differ only by the overall scale factor $\cosh^2 t_E$, the same is true of their quotients.  The spatial slice at $t=0$ (equivalently, $t_E = 0$) is thus conformal to the geometry at $t_E = -\infty$.  This is half of the Euclidean boundary, with the other half being $t_E = +\infty$. We may therefore write $|TFD \rangle$ as given by the CFT path integral over the Riemann surface defined by the $t=0$ slice of the corresponding BTZ geometry.

This final conclusion can be extended to a much larger class of states.  Any Riemann surface $\Sigma$ with $n$ boundaries can be written as a quotient of $H^2$ by some discrete subgroup $\Gamma_\Sigma$ of $SL(2,{\mathbb R})$.  We may use \eqref{eq:hypslice} to construct a corresponding quotient of Euclidean AdS${}_3$, with $\Sigma$ conformal to both the slices at $t=0$ and $t_E = -\infty$.  So long as this saddle point dominates the bulk path integral with boundary conditions defined by the $t_E = -\infty$ slice, to good approximation the corresponding Lorentz-signature bulk solution -- given by substituting $t_E = -i t$ into \eqref{eq:hypslice} -- is dual to the CFT state defined by the path integral over the slice at $t=0$.  For notational simplicity we identify $\Sigma$ with this slice below and write the CFT state as $|\Sigma\rangle.$ These quotients of AdS${}_3$ were first considered in \cite{Aminneborg:1997pz}, and the holographic relation to $|\Sigma \rangle$ was introduced in  \cite{Krasnov:2000zq,Krasnov:2003ye,Skenderis:2009ju}. An exploration of the entanglement properties of these states was initiated in \cite{Balasubramanian:2014hda}.

The Lorentz-signature solutions describe wormholes connecting $n$ asymptotically-AdS$_3$ boundaries.  By topological censorship \cite{Friedman:1993ty,Galloway:1999bp}, each boundary is associated with a distinct event horizon. A special property of AdS$_3$ vacuum solutions is that the geometry outside each event horizon is precisely that external to some BTZ black hole.  This allows us to define a useful ``round'' conformal frame, in which the usual rotational symmetry of this BTZ region is a symmetry of the boundary. That is, for each of these exterior regions there is a coordinate $\phi_i$ such that $\partial_{\phi_i}$ is an exact rotational Killing field in the region outside the horizon (and in fact in an open neighbourhood in the interior of the horizon as well). The round conformal frame is the one in which the CFT lives on a spacetime with standard cylinder metric
\begin{equation}
ds^2 = - dt^2 + d\phi_{i}^2
\end{equation}
with $\phi\sim\phi+2\pi$.
In addition, the BTZ exterior implies that the bifurcation surface of each horizon -- where the future and past horizons meet -- is a geodesic in the $t=0$ surface. The key novelty in the $n >2$ cases is the existence of a ``causal shadow'' region in between these horizons.

\begin{figure}
\centering
\includegraphics[keepaspectratio,width=0.5\linewidth]{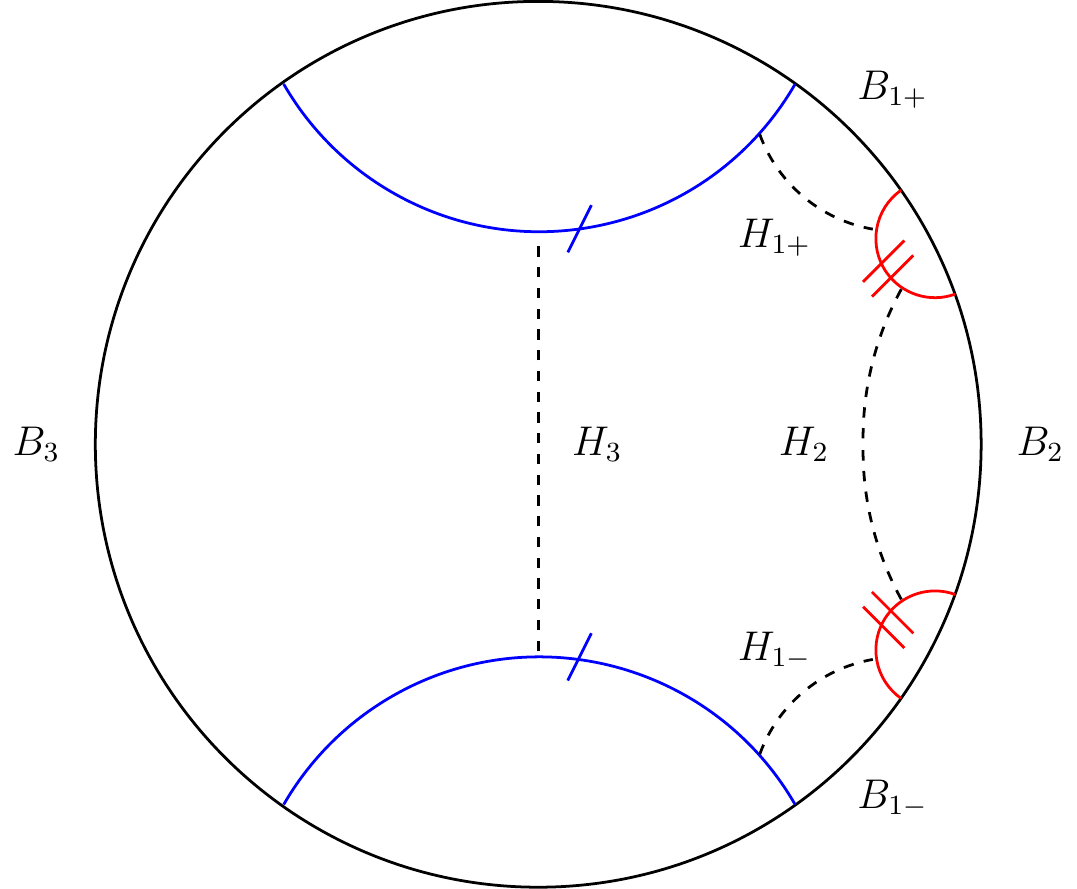}
\caption{The surface $\Sigma$ as a quotient of the Poincar\'e disc for $n=3$. The pairs of labeled geodesics (blue and red in colour version) are identified by the action of $\Gamma$. The region of the Poincar\'e disc bounded by these geodesics provides a fundamental domain for the quotient. $B_3$, $B_2$ and $B_1 = B_{1+} \cup B_{1-}$ become the desired three circular boundaries. There are corresponding minimal closed geodesics $H_3$, $H_2$ and $H_1  = H_{1+}\cup H_{1-}$. The lengths $L_a$ of these geodesics fully characterize the geometry of $\Sigma$.}
\label{pants}
\end{figure}

 Our ideas will apply to a codimension one limit in the moduli space of such Riemann surfaces for any $n$, but for simplicity we will focus our discussion on the case where $\Sigma$ is an orientable surface with three boundaries and no handles. Such surfaces are topologically the same as a pair of pants. This is the simplest non-trivial example, and is also a primitive building block for constructing other cases, since a general orientable Riemann surface can be constructed by  sewing together pairs of pants. The relevant quotient of $H^2$ is depicted in figure \ref{pants}. The moduli space of pair-of-pants Riemann surfaces can be parametrized by the lengths $L_a$ ($a=1,2,3$) of the three horizons, which as usual we take to be measured in units with $\ell_{AdS} =1$.  Without loss of generality we take $L_3 \geq L_1, L_2$. The causal shadow is the region in between these geodesics.

 Properties of such states were explored in \cite{Balasubramanian:2014hda}, with most emphasis on the so-called puncture limit $L_a \ll 1$. In particular, \cite{Balasubramanian:2014hda} showed that in this limit  $\Sigma$ is conformal to the Riemann sphere with small holes removed around $n$ points, and hence $|\Sigma \rangle$ can be related to an $n$-point function in the CFT. For the three-boundary case, the state was determined up to some constant factors to be, in the round conformal frame,
\be \label{gen3stateplane}
|\Sigma \rangle = \sum_{ijk}  C_{ijk} e^{-\frac{1}{2}\tilde \beta_1 H_1}e^{-\frac{1}{2}\tilde \beta_2 H_2}e^{-\frac{1}{2}\tilde \beta_3 H_3}    |i\rangle_1 |j\rangle_2 |k\rangle_3,
\ee
where
\be
\tilde \beta_a = \beta_a - 2 \ln r_d  - 2 \ln 3,
\ee
and $C_{ijk}$ are the three-point OPE coefficients,  $\beta_a = \frac{4\pi^2}{L_a}$ is the inverse temperature of the BTZ geometry associated with the region near the $a$'th boundary, and $r_d$ is an undetermined constant independent of the moduli. The rather explicit expression \eqref{gen3stateplane} exhibits both dependence on the structure of the CFT and Boltzmann-like suppression factors similar to the thermofield double state.

\section{Geometry of $\Sigma$ in the high temperature limit}
\label{sec:LargeL}

Our current aim is to elucidate the structure of $|\Sigma \rangle$ in the limit $L_a \rightarrow \infty$ with fixed ratios $L_a/L_b$. This is the opposite of the limit emphasized in \cite{Balasubramanian:2014hda,Pastawski:2015qua}. We assume $L_3 \geq  L_1, L_2$, so the ratios $L_1/ L_3, L_2/L_3$ take values in $(0,1]$. In this limit, the geometry of $\Sigma$ again simplifies. The essential point is that the causal shadow region will play a relatively unimportant role. We will focus on the pair of pants case, but the discussion is easily extended to arbitrary Riemann surfaces. We comment on this extension in section \ref{disc}.

Our limit can be characterised as a high temperature limit, in the sense that the BTZ horizon in each of the exterior regions becomes large compared to the AdS scale (as for a high $T$ BTZ black hole).  But we note that the restriction of the state $|\Sigma \rangle$ to a single boundary is not necessarily even approximately thermal: as discussed in \cite{Balasubramanian:2014hda}, when one $L_a$ is larger than the sum of the other ones, reduced density matrix in that exterior region has much less entropy than the thermal value at the same energy.

\begin{figure}
\includegraphics[scale=1]{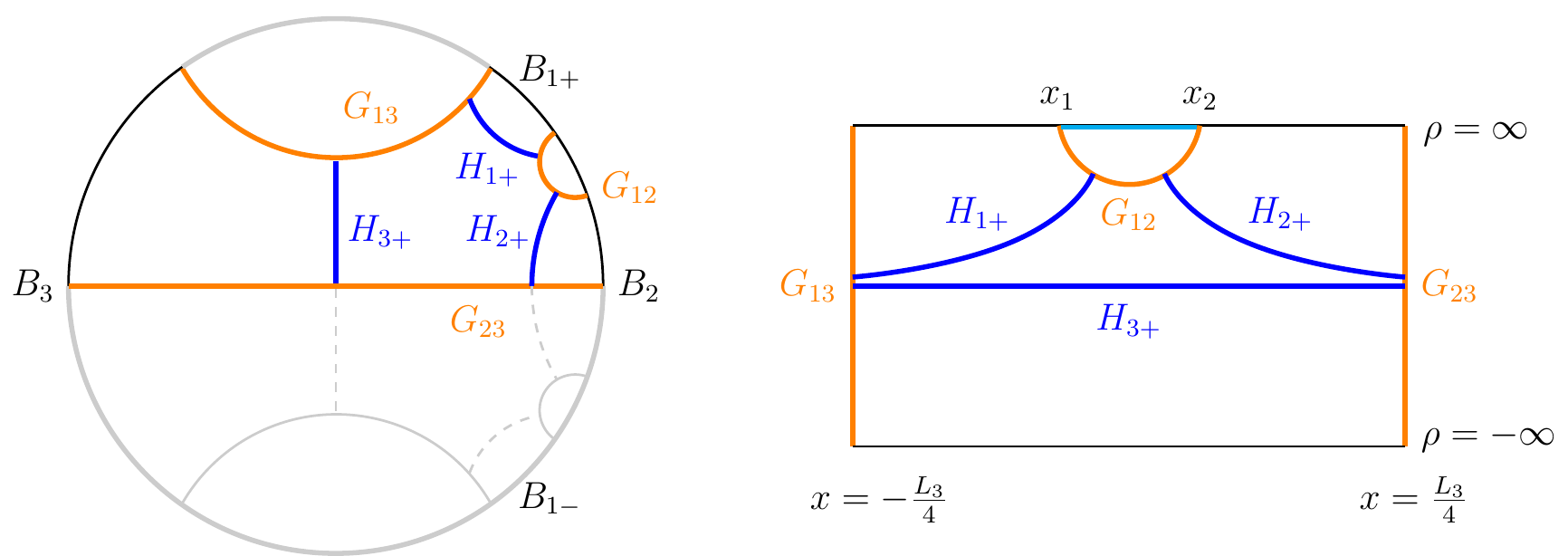}
\caption{The region $\Sigma_{+}$ bounded by the geodesics $G_{ab}$, half each of $B_2,B_3$, and $B_{1+}$ shown in the Poincar\'e disc (left) and the BTZ frame strip (right). The BTZ presentation is chosen to place the half-horizon $H_{3+}$ along the BTZ horizon. The geodesics $G_{13}$, $G_{23}$ are respectively the lines $x=- \frac{L_3}{4}$, $x=\frac{L_3}{4}$.  In contrast, $G_{12}$ lies in the upper half of the strip; its endpoints have $x=x_1,x_2$ with $\rho=+\infty$.  Half each of $B_1, B_2$ is mapped respectively to the line segments $ x \in [-\frac{L_3}{4}, x_1]$, $ x \in [x_2,\frac{L_3}{4}]$ at $\rho=\infty$ respectively, whilst half of $B_3$ is mapped to $\rho=-\infty$. The corresponding $\Sigma_-$ is the symmetric region below $G_{23}$ in the Poinar\'e disk (left) and has an identical representation in the BTZ strip. }
\label{fig:maptostrip}
\end{figure}

To understand the geometry of $\Sigma$ in our limit, it is useful to introduce a different presentation using two patches with BTZ coordinates on each\footnote{In the actual history of our project, this description was also inspired by computing mutual information on pairs of pants with large $L_a$ using the technology of \cite{Maxfield:2014kra}.}. We split figure \ref{pants} in half along the horizontal geodesic (not drawn explicitly) joining boundaries $B_1$ and $B_2$.  This divides $\Sigma$ into two identical regions $\Sigma_{\pm}$, each containing half of each horizon $H_a$.  The surface $\Sigma$ is then recovered by gluing together $\Sigma_{\pm}$ along three geodesics, the two identified geodesics in figure \ref{pants} and the new split. We label these geodesics $G_{ab} = G_{ba}$ with $a\neq b$ labelling the boundaries they run between; see figure \ref{fig:maptostrip} (left).  They can be described more formally as the fixed points of a $\mathbb{Z}_2$ isometry of $\Sigma$, which acts as a reflection $\phi \rightarrow 2 \pi - \phi$ in the round conformal frame on each of the boundaries (with appropriate choices of the origin $\phi=0$ on each boundary). The event horizon $H_a$ of boundary $a$ is the unique geodesic that runs orthogonally between the two geodesics $G_{ab}$, $G_{ac}$ $(b\neq c)$ that end on boundary $a$.  Our partition of $\Sigma$ into $\Sigma_\pm$ also breaks each horizon $H_a$ into two pieces $H_{a\pm}$.

\begin{figure}
\centering
\begin{subfigure}[t]{0.7\textwidth}
\centering
\hspace*{-15pt}
\includegraphics[scale=0.83]{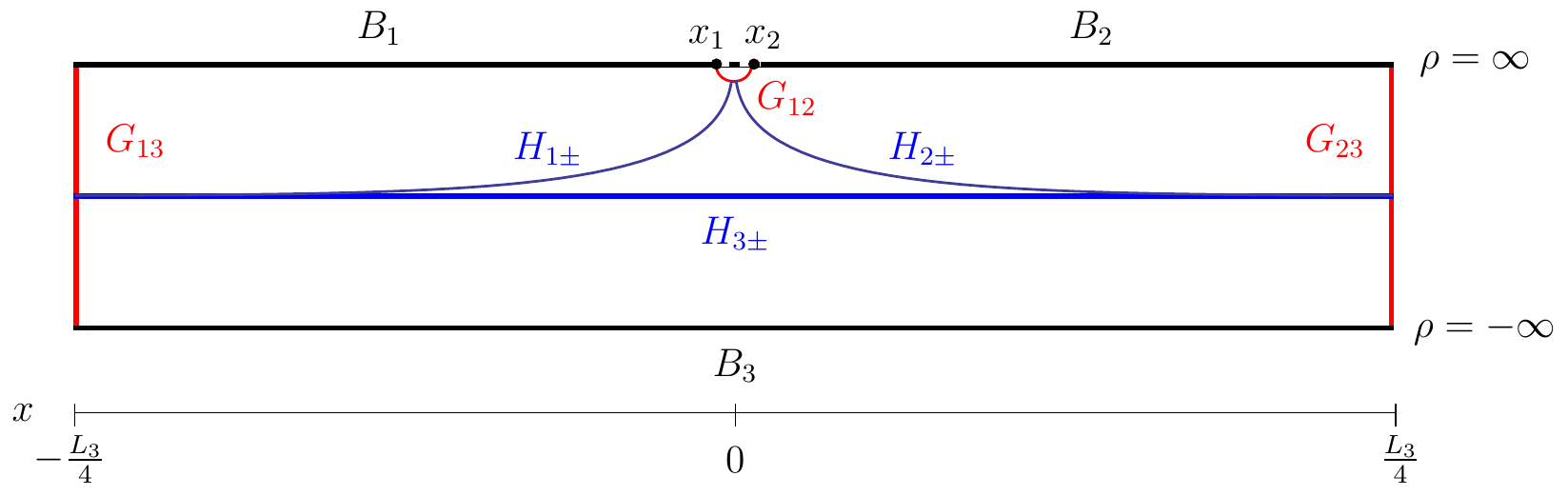}
\caption{$L_1 + L_2 > L_3$}
\end{subfigure}\\
\begin{subfigure}[t]{0.7\textwidth}
\centering
\includegraphics[width=\textwidth]{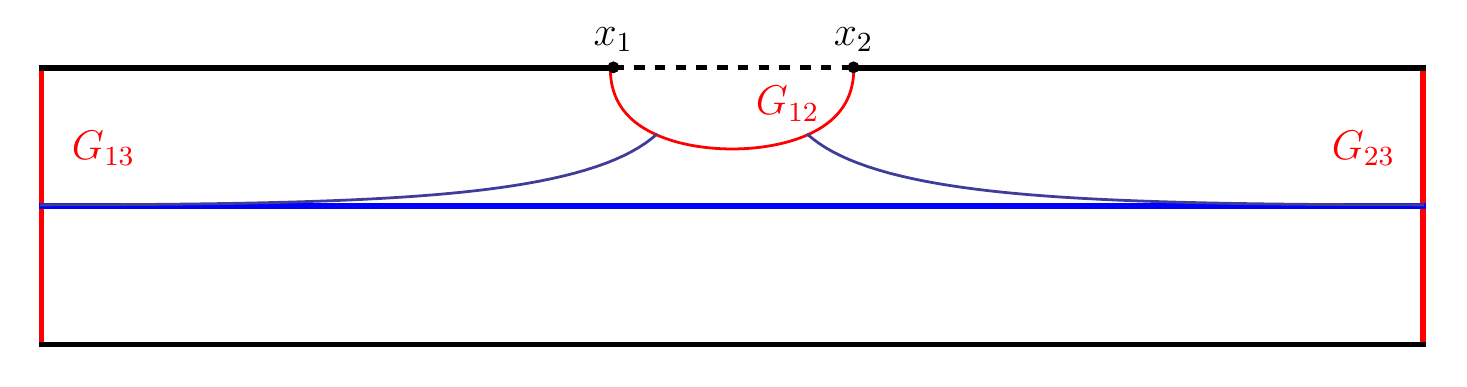}
\caption{$L_1 + L_2 \sim L_3$}
\end{subfigure}\\
\begin{subfigure}[t]{0.7\textwidth}
\centering
\includegraphics[width=\textwidth]{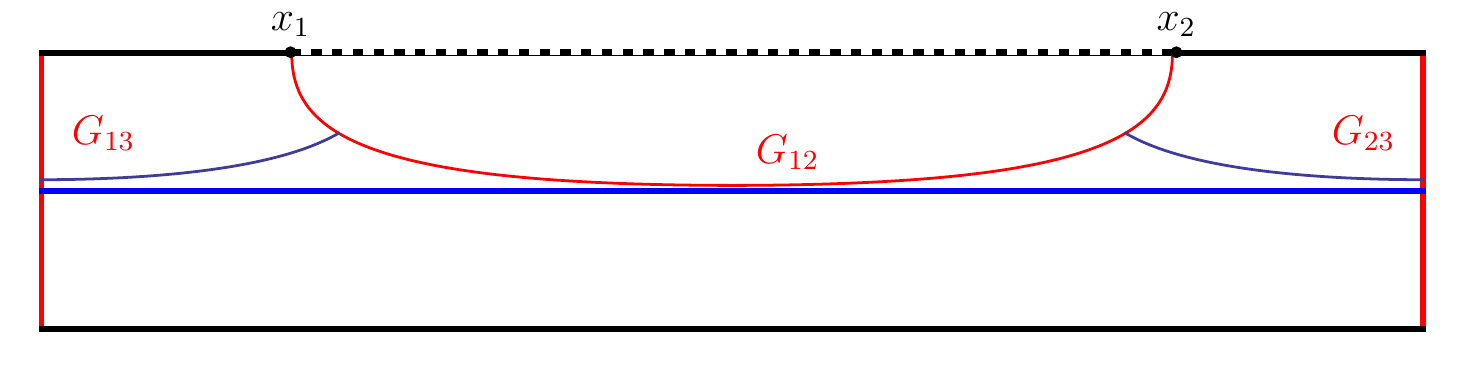}
\caption{$L_3 > L_1 + L_2$ }
\end{subfigure}
\caption{Half of the pair-of-pants (either $\Sigma_+$ or $\Sigma_-$) described as a region in planar BTZ.  Three examples are shown representing distinct regions of moduli space: $L_1 + L_2 > L_3$ (top), $L_1 + L_2 \sim L_3$ (middle), $L_3 > L_1 + L_2$ (bottom).}
\label{fig:BTZrep}
\end{figure}

It is useful to describe $\Sigma_\pm$ in planar BTZ coordinates.\footnote{By planar BTZ we mean the usual BTZ coordinates with no identification on the spatial coordinate on the boundary; this provides a coordinate system on the whole of $H^2$, thought of as the $t=0$ surface in AdS$_3$. Since our $\Sigma_\pm$ are subregions of $H^2$, they can be conveniently described in these coordinates.} Consider for definiteness $\Sigma_+$. We choose the BTZ coordinates to be
\begin{equation}
\label{eq:BTZ}
\frac{ds^2_{BTZ}}{\ell_{AdS}^2} = \frac{d\rho^2}{\rho^2 +1} + (\rho^2+1) dx^2,
\end{equation}
with $\rho \in (-\infty, \infty)$.  Thus our reference BTZ solution has inverse temperature $2\pi$.  Without loss of generality, we take $L_3 \ge L_2 \ge L_1$ and orient $\Sigma_+$ such that the portion of $H_3$ in $\Sigma_+$ lies along the horizon at $\rho=0$,  and the boundary $B_3$ lies at $\rho = -\infty$, in both cases for $x \in [-L_3/4, L_3/4]$. Since the geodesics $G_{13}$ and $G_{23}$ intersect $H_3$ orthogonally, they will lie at $x = -L_3/4$ and $x = L_3/4$ in these coordinates. The other two boundaries $B_1$ and $B_2$ lie at $\rho = \infty$, for $x \in [-L_3/4, x_1]$ and $x \in [x_2, L_3/4]$ (with $x_1 < x_2$), and the remaining geodesic $G_{12}$ runs between these points $x_1, x_2$. The portions of $H_1$, $H_2$ in $\Sigma_\pm$ are the geodesics running from the edges of the strip to meet $G_{12}$ orthogonally. These coordinates are illustrated in figure \ref{fig:maptostrip}.

The half-surface $\Sigma_+$ is thus a strip $x \in  [-L_3/4, L_3/4]$ in the planar BTZ coordinates, with a bite cut out of the middle above $G_{12}$. It is important to emphasize that the boundaries at $-L_3/4, L_3/4$ are not identified with each other; instead they and $G_{12}$ are identified with the corresponding boundaries in a second copy of this region.

As we verify in appendix \ref{app:H1H2}, varying the endpoints $x_1, x_2$ of $B_1,B_2$ generates all possible lengths $L_1,L_2$ for the remaining horizons $H_1,H_2$ and the map $(x_1,x_2) \mapsto (L_1,L_2)$
is both smooth and one-to-one. When we take the limit of large $L_a$ (at fixed ratios), the results simplify, with a form that depends on the relative lengths. For $L_3- (L_1+L_2) \gg 1$,
\begin{equation}\label{eq:L1L2}
	x_1 \sim \frac{L_1}{2}-\frac{L_3}{4}- \log 2, \; x_2 \sim \frac{L_3}{4}-\frac{L_2}{2}+ \log 2,
\end{equation}
where the tildes ($\sim$) represent agreement up to exponentially small corrections.  Note that up to a fixed order-one offset,  the endpoints are respectively $L_1/2$ and $L_2/2$ from the ends of the strip.
In the complementary case $L_3- (L_1+L_2) \ll 1$, we find instead
\begin{equation}
	\frac{x_1+x_2}{2} \sim \frac{L_1-L_2}{4}, \ \  \ \frac{x_2-x_1}{2} = \exp\left(-\frac{L_1+L_2-L_3}{4}\right).
\end{equation}

In our BTZ presentation, the long length of $H_3$ corresponds directly to the large width of the strip. The horizons $H_1, H_2$ are also long as a result of extending over a large coordinate distance in the $x$ direction and possibly also from extending out towards the boundary of the strip at large $\rho$.  If both of them together are shorter than $H_3$ ($L_1 + L_2 \leq L_3$), they terminate on $G_{12}$ in the interior of the strip, staying within an order one distance from the horizon $H_3$ at $\rho=0$ along their whole length, as in the last panel of figure \ref{fig:BTZrep}. When the sum is larger (which includes the case where the three lengths are equal), the length of the interval $x_2-x_1$ is exponentially short,  hence $H_1,H_2$ meet $G_{12}$ at large $\rho$, as in the first panel of figure \ref{fig:BTZrep}.  In both cases, $H_1$ and $H_2$ approach $H_3$ exponentially for $|x-x_{1,2}| \gg 1$.

The contributions to $L_1,L_2$ from the width of the strip or from $H_1$ and $H_2$ running to large $\rho$ look different, but we should remember that $\Sigma_+$ treats the three horizons symmetrically, so this is just an artifact of our choice of coordinates. The symmetry can be made manifest in an appropriate Poincar\'e disk representation; see e.g. figure 8 of \cite{Balasubramanian:2014hda}.  The relationship between any pair of horizons is thus much the same; consider for example $H_1$ and $H_3$. We can see explicitly from the calculation in appendix \ref{app:H1H2} that the minimal distance between them is exponentially small, and that they remain exponentially close over a large region. Thus, the area of the causal shadow region remains finite even as their length becomes large.

In fact, since the boundaries of the causal shadow are closed geodesics (and thus have vanishing extrinsic curvature), the Gauss-Bonnet theorem requires this area $A_{CS}$ to be independent of the moduli $L_a$ (for any fixed genus $g$ and number of boundaries $n$).  For the pair of pants we find $A_{CS} = 2 \pi$; more generally $A_{CS} = 2(n-2 + 2g)\pi$. As we will see in the next section, this implies that the causal shadow region plays little role in the path integral construction of the CFT state $|\Sigma \rangle$.

In the case where $L_3 - (L_1 + L_2) \gg 1$, the endpoints of the geodesic $G_{12}$ 
are far apart in coordinate distance, and it will also be exponentially close to $H_3$ over most of its length. When we glue $\Sigma_+$ and $\Sigma_-$ to form $\Sigma$, the section of $H_3$ that is close to $G_{12}$ will lie close to the corresponding section of $H_3$ in $\Sigma_-$, as in figure \ref{fig:sew} (right).  All remaining cases with $L_3 \ge L_2 \ge L_1$ are intermediate between the two just described.

\section{The CFT state at large $L_a$}
\label{sec:CFT}

Let us now consider the implications of the above results on the structure of $\Sigma$ for the CFT state $|\Sigma \rangle$. In this section we will argue for large $L_a$ that $|\Sigma\rangle$  will be described to exponential accuracy by figure \ref{fig:sew} (right).  In particular, when restricted to appropriate regions it agrees to exponential accuracy with the corresponding restriction of a thermofield double state $|TFD\rangle$.  We also show, under the assumption that non-handlebody contributions can be ignored, that the Euclidean bulk geometry \eqref{eq:hypslice} dominates the bulk path integral defined by using $\Sigma$ as the conformal boundary.  It follows that, to exponential accuracy, our bulk pair-of-pants wormhole is dual to the state described by figure \ref{fig:sew} (right).

\begin{figure}
\centering
\begin{subfigure}[t]{0.7\textwidth}
\centering
\includegraphics[scale=0.8]{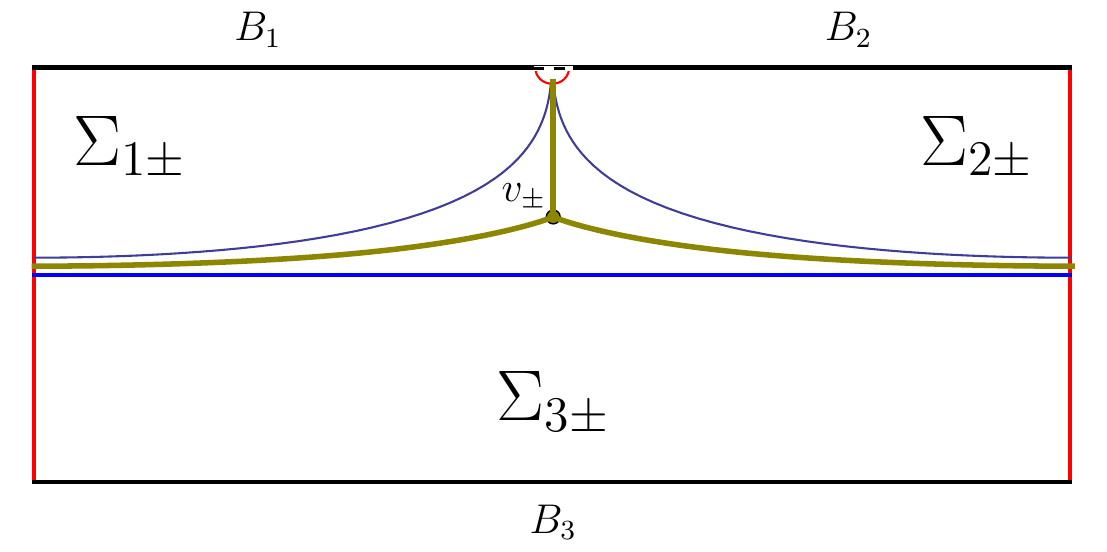}
\end{subfigure}
\begin{subfigure}[t]{0.7\textwidth}
\centering
\includegraphics[scale=0.8]{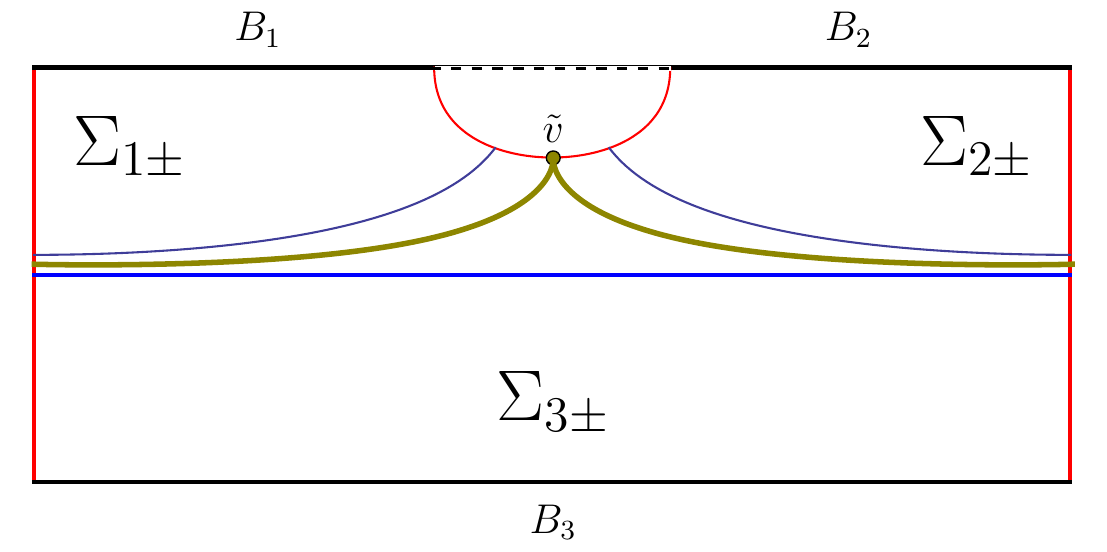}
\end{subfigure}
\begin{subfigure}[t]{0.7\textwidth}
\centering
\includegraphics[scale=0.8]{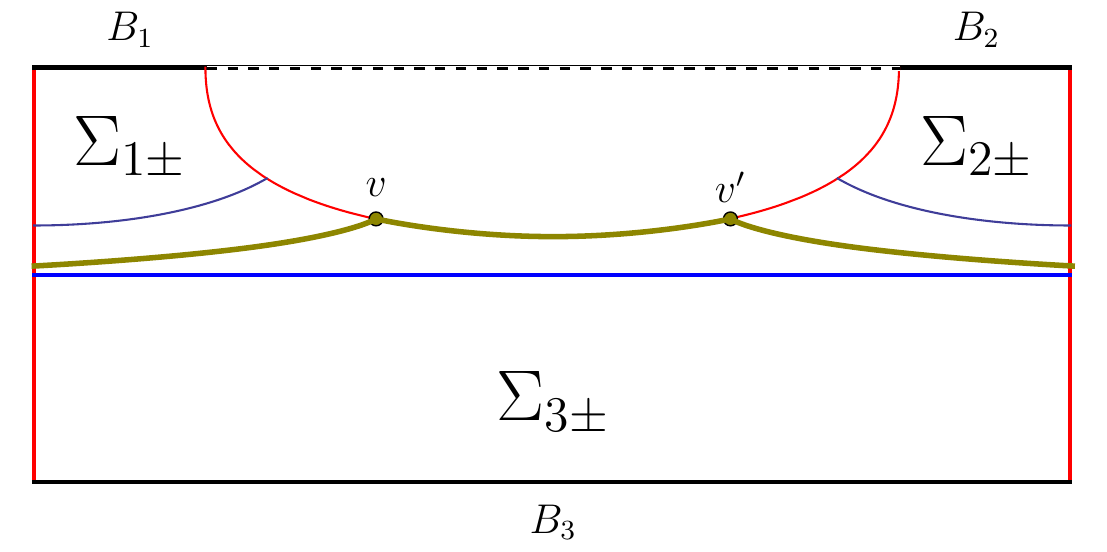}
\end{subfigure}
\caption{The decomposition of $\Sigma_\pm$ into pieces $\Sigma_{1\pm},\Sigma_{1\pm},\Sigma_{1\pm}$.  We then glue $\Sigma_{a+}$ to $\Sigma_{a-}$ along the relevant $G_{ab}$ to make pieces $\Sigma_a$ conformal to cylinders with $\Sigma_a$ containing all of boundary $a$.  The decomposition is determined by a graph.  In cases (a) and (c), the graph has two trivalent vertices.  In case (a) each piece $\Sigma_\pm$ contains one vertex $v_\pm$.  The 3 edges of the graph each connect $v_+$ to $v_-$ running between two distinct horizons $H_a, H_b$ for $a \neq b$.      In case (c) both vertices $v, v'$ lie on the cut along $G_{12}$, as does the edge that connects them.  The other two edges are loops connecting $v$ to $v$ or $v'$ to $v'$.  One lies between $H_1$ and $H_3$ while the other lies between $H_2$ and $H_3$. Case (b) represents a degenerate limit interpolating between the two in which we choose to fuse the two vertices into a single 4-valent vertex lying on $G_{12}$.  The graph has two edges, each of which are loops.  One lies between $H_1$ and $H_3$ while the other lies between $H_2$ and $H_3$.  In the regions far from the vertices the pieces $\Sigma_{1,2,3}$ defined by cutting $\Sigma$ along the edges of the appropriate graph differ from the cylinders defined by the regions outside horizons $H_{1,2,3}$  only by exponentially small amounts.}
\label{fig:PIdecomp}
\end{figure}

Recall that $|\Sigma \rangle$ is defined by the CFT path integral over $\Sigma$.  We will use the BTZ representation of $\Sigma_\pm$ associated with figure \ref{fig:BTZrep} to break $\Sigma$ into three pieces $\Sigma_{1,2,3}$ that are topologically cylinders, corresponding to figure \ref{fig:sew} (left).  Each piece $\Sigma_{a}$ will contain the entire pair-of-pants boundary $B_a$, but no portion of the boundaries $B_b$ for $b\neq a$.   The decomposition is defined by drawing a graph in $\Sigma$ as shown in figure \ref{fig:PIdecomp}.  As noted in the figure caption, far from the vertices the piece $\Sigma_a$ nearly coincides with the region $E_a$ of $\Sigma$ exterior to $H_a$ (i.e., lying between $H_a$ and boundary $a$).  The geometry in this latter region is just that of the appropriate BTZ solution outside the horizon and is conformal to a round (rotationally-invariant) cylinder.

We wish to regard both $E_a$ and $\Sigma_a$ as path integrals constructing states $|E_a\rangle$, $|\Sigma_a \rangle$, each of which is defined on two copies of our CFT (on the outer and inner boundaries of $E_a$ or $\Sigma_a$ respectively). Indeed, we may write $|\Sigma_a \rangle = \hat S_a |E_a\rangle$ where $\hat S_a$ is the operator defined by the path integral over the causal shadow region $S_a=\Sigma_a/E_a$ in $\Sigma_a$ beyond the horizon $H_a$.  We specify the conformal frames of all states by again taking $\phi_a$ to define the standard angle on the CFT cylinder; this involves a natural extension of $\phi_a$ through the causal shadow $S_a$.   The region $S_a$ is topologically an annulus and so can be conformally transformed to a cylinder.  But $S_a$ is exponentially thin over most of its circumference; indeed, setting $a=3$ (so that we may replace $\phi_3$ by $2\pi/L_3$ times the BTZ $x$) and multiplying the BTZ metric \eqref{eq:BTZ} by $\ell_{AdS}^{-2}(1 +\rho^2)^{-1}$ gives a metric
\begin{equation}
\label{eq:expcyl}
ds^2_{S_3} = dx^2 + dy^2,
\end{equation}
where $y = \tan^{-1} \rho$ ranges over $[0, f(x)]$ with $f(x)$ is exponentially small far from the vertices of our graph.
Introducing $\tilde y = y/f$ along with $\tilde x$ such that $d\tilde x = dx/f$, and multiplying \eqref{eq:expcyl} by $f^{-2}$ gives a metric
\begin{equation}
\label{eq:expcyl2}
f^{-2} ds^2_{S_3} = d\tilde x^2 + (d\tilde y + f' \tilde y d\tilde x)^2,
\end{equation}
where $f' = df/dx$, on cylinder of unit height $\tilde y \in [0,1]$ but with exponentially large circumference.  The metric is not flat, though it differs from the standard cartesian flat metric $d\tilde x^2 + d \tilde y^2$ only by exponentially small corrections proportional to powers of $f'$.    It follows that there is a further conformal transformation to a metric cylinder of unit height -- and with exponentially small difference in circumference from the range of $\tilde x$ -- whose action on the region far from the vertices is exponentially close to the identity map\footnote{Here we use the fact that conformal transformations satisfy an elliptic equation with a Green's function that decays exponentially along a strip.  We expect that similar arguments are common in the literature, but for a specific example the interested reader interested in details may consult for comparison e.g. section 3.1.1 of \cite{Balasubramanian:2014hda}.}.

Rescaling this cylinder to one of circumference $2\pi$ allows us to write the path integral over $S_a$ in terms of the operator $e^{-\beta H}$ with exponentially small $\beta$.  Up to exponentially small corrections, this operator acts as the identity with respect to degrees of freedom associated with spatial regions of order-one size as measured by the original spatial coordinate $\phi_a$.  So far from the vertices we may identify $|E_a\rangle$ and $|\Sigma_a \rangle$ with exponential accuracy.

On the other hand, the exterior region $E_a$ is half of the BTZ $t=0$ surface, so $|E_a \rangle$ is given by a path integral over a cylinder of length $\beta/4$, so it is a copy $|TFD \rangle_a$ of \eqref{eq:TFD} at twice the temperature of the associated bulk horizon. Thus $|\Sigma_a \rangle$ is, up to exponentially small corrections, a thermofield double state. Recall that for large $L_a$ such $|TFD\rangle_a$ are described by a tensor network of the form shown in \cref{fig:TFDnet}.

It remains only to sew the $|\Sigma_a\rangle = |TFD\rangle_a$ together into $|\Sigma\rangle$.
The sewing procedure is defined by the way the path integrals $\Sigma_{1,2,3}$ combine to form $\Sigma$. Away from the vertices of the graph, this identifies the horizons in $E_{1,2,3}$: for $L_1 + L_2 > L_3$, parts of $H_3$ are identified with each of $H_1$ and $H_2$, and the remaining parts of $H_1$ and $H_2$ are identified with each other. For $L_1 + L_2 < L_3$, $H_1$ and $H_2$ are each entirely identified with corresponding parts of $H_3$, and the remaining regions of $H_3$ along $G_{12}$ are identified with each other. Since the sewing operation on path integrals coincides with the sewing operation on tensor networks -- one simply sets all arguments equal along the seam and integrates over allowed values\footnote{Here for simplicity we again make use of the time-reversal symmetry mentioned in \cref{review} to turn bra-vectors into ket-vectors.}-- this implies that the state $|\Sigma \rangle$ is given to exponential accuracy by \cref{fig:sew} (right), with the top picture relevant for $L_1 + L_2 > L_3$ and the bottom picture relevant for $L_1 + L_2 < L_3$.

Finally, we also wanted to see that $|\Sigma \rangle$ is dual to our bulk geometry with moment of time symmetry $\Sigma$.  As in \cite{Balasubramanian:2014hda} we assume that the dominant saddle of the associated bulk integral is a handlebody. The other possible bulk saddles discussed in \cite{Balasubramanian:2014hda} correspond to disconnected Lorentzian geometries.  It is natural to expect this saddle to dominate at large temperatures, by analogy to the familiar result for $|TFD\rangle$ that disconnected solutions dominate only at low temperatures. But one can now make a further argument based on entanglement.  If the HRT proposal is correct, and in particular if entanglement is associated with extremal surfaces in the real Lorentz-signature geometry, the disconnected geometries cannot reproduce the entanglement structure of figure \ref{fig:sew} (right), which involves entanglement between the different boundaries at leading order in the central charge.  It would be interesting to verify this conclusion by direct computation of the Euclidean actions, as it would serve as a check on HRT.

\section{Holographic entanglement calculations}
\label{holo}

The previous section used the CFT path integral to show that the CFT state $|\Sigma \rangle$ is given by figure \ref{fig:sew} (right), so that the state has local bipartite entanglement with the same local structure as the thermofield double state. In this section we will buttress that argument by showing that our picture of the geometry of $\Sigma$, now thought of as the $t=0$ surface in the bulk spacetime, gives consistent results for entanglement from holographic Ryu-Takayanagi calculations. Indeed, in the history of our project we originally discovered that the state had this simple bipartite structure by performing these holographic calculations explicitly. We consider the entanglement for a region in boundary $3$, since our coordinates are adapted to this boundary, but by symmetry similar results apply in the other cases.

Consider first a region in boundary $3$ where the horizon $H_3$ is exponentially close to either $H_1$ or $H_2$, that is $x_1 - x \gg 1$ or $x - x_2 \gg 1$. In the exterior region $E_3$, the planar BTZ coordinate $x$ is identical (up to a scale and a shift of origin) to the round conformal frame coordinate $\phi_3$ defined in section \ref{review}: $\phi_3 = \frac{2\pi}{L_3} x$. In the other exterior region, at similar $x$ but outside $H_1,H_2$, because the horizon $H_{1,2}$ is exponentially close to $H_3$, the planar BTZ coordinate $x$ agrees with $\phi_{1,2}$ (up to a scale and a shift of origin) up to exponentially small corrections. We can take for example $\phi_1 = \frac{2\pi}{L_1} x$.  This is manifestly true near $\rho = 0$ (see \eqref{eq:BTZ}) and continues to hold at large $\rho$ due to the properties of geodesics in hyperbolic geometry\footnote{Two geodesics on $H^2$ fired at slightly different angles from the same point will diverge exponentially as measured by the proper distance separating them as the curves approach the boundary. So curves of constant $\phi_{1,2}$ and $x$ that meet at the  horizon also diverge in a similar manner near the  boundary.   But two geodesics fired orthogonally from different points $x'$, $x''$ of the  horizon again diverge exponentially at precisely the same rate.  So a curve of constant $\phi_{1,2}$ that meets the  horizon at $x'$ with $|x' - x_{1,2}| \gg 1$ will meet the boundary at some $x''$ with $|x''-x'|$ still exponentially small.}.  So for $x_1-x \gg 1$, $x-x_2 \gg 1$  we may take $x$ to define the round conformal frame on all three boundaries up to exponentially small corrections.

The above relations allow us to easily map those geodesics involved in any HRT calculation of the mutual information between subregions of boundaries $1$ and $3$ (or $2$ and $3$) that lie far from $x_1,x_2$ to geodesics in BTZ. The BTZ calculation was studied in \cite{Morrison:2012iz}, who found that for regions much larger than the thermal scale, the mutual information is simply proportional to the size of the overlap between the two regions. The overlap is maximal when the two regions are directly opposite each other, in which case the high-temperature result (3.27) of \cite{Morrison:2012iz} becomes
\begin{equation} \label{Ilin}
I(A:B) = S(A) + S(B)  - S(A \cup B) = \frac{L}{4G} \frac{( \Delta \phi + 2\pi  -  (2\pi - \Delta \phi) )}{2\pi} = \frac{L}{2G}  \frac{\Delta \phi}{2\pi} + {\cal O}(L^0).
\end{equation}
Applying appropriate scalings to \eqref{Ilin}, the mutual information between corresponding regions of boundaries $1$ and $3$ with  $x_1-x \gg 1$ is
\begin{equation}
\label{eq:MIhighT}
I(A:B) = \frac{1}{2G} \Delta x+ O(1) =  \frac{L_1}{2G} \frac{\Delta \phi_1}{2\pi} + O(1) =  \frac{L_3}{2G} \frac{\Delta \phi_3}{2\pi} + O(1).
\end{equation}
In addition,  since the region of boundary $1$ with $x_1-x \gg 1$ is well-separated in the bulk from the region of boundary $2$ with $x-x_2 \gg 1$, it also follows that these two regions share no mutual information. The situation is exactly similar for the region in boundary $3$ with $x-x_2 \gg 1$, which has a mutual information of the same form with a region in boundary $2$.

If $|x_2 - x_1| \ll 1$, there are large parts of $H_1, H_2$ that are far from $H_3$, and so have yet to be described.  This indicates that there are large intervals of $\phi_{1,2}$ along boundaries $1$ and $2$ with $x$-values close to the endpoints $x_1,x_2$.  But it also implies a large conformal transformation between the round conformal frame for $B_1$, $B_2$ and the planar BTZ coordinate $x$.  As a result, the renormalized length of any geodesic connecting boundary $3$ to these regions of boundaries $1,2$ is very long and HRT calculations give no mutual information between boundary $3$ and these regions.

For $|x_2 - x_1| \ll 1$, the above results describe the entanglement properties of boundary $3$, with the exception of a region with length of order the thermal length scale (which in the planar BTZ coordinates is of order the AdS scale) near $x_1,x_2$.  So away from the vertices the entanglement structure obtained from bulk calculations corresponds precisely with that predicted by the state pictured in the top panel of figure \ref{fig:sew}, given by sewing together thermofield double states.

We now turn to the complementary case $|x_2 - x_1| \gg 1$. There is then a large region of boundary $3$ not covered by the regions $x_1-x \gg 1$, $x-x_2 \gg 1$ studied above.  But across the region satisfying both $x - x_1 \gg 1$ and $x_2 - x \gg 1$, the geodesic $G_{12}$ lies exponentially close to $H_3$.  So sewing together $\Sigma_+$ and $\Sigma_-$ in this region is well-approximated by simply gluing to each other the boundaries of $E_3$ (the region outside $H_3$; the lower half of each diagram in figure \ref{fig:BTZrep}) along $H_{3\pm}$; i.e., the result is well approximated by the region of the two-sided BTZ geometry with $x_1 < x < x_2$.

Note that the two asymptotic boundaries of this new BTZ geometry are identified with different regions of boundary $3$ coming respectively from $\Sigma_+$ and $\Sigma_-$.  In particular, since our slicing of the pair of pants into $\Sigma_\pm$ was performed using the $\mathbb{Z}_2$ reflection symmetry, we see that a given value of $x$ with $x_1 < x < x_2$ corresponds both to some point $\phi_3$ and also to $2 \pi - \phi_3$ in terms of the usual coordinate on boundary $3$ that defines the round conformal frame. Thus, in this case the mutual information of a region in boundary $3$ with the corresponding region on the opposite side in boundary $3$ will be as given in \eqref{eq:MIhighT}, supporting the local thermofield-double like entanglement between the two pieces of this boundary as indicated in the bottom panel of \cref{fig:sew}.

\section{Finite size corrections}\label{finite}

We have shown that $|\Sigma \rangle$ has a simple structure in the high temperature limit.  To use this as a systematic approximation to the state corresponding to finite-size wormholes, it is interesting to investigate finite-temperature corrections to this. In this section we will consider this first for the simple two-boundary case and then for three boundaries.

\subsection{Two boundaries}
\label{two}

In the two-boundary case, we want to understand and characterise the departure from the trivial network pictured in figure \ref{fig:TFDnet}. The departure will be significant when we consider small regions, of order the thermal scale or smaller. For simplicity, we diagnose this by considering the mutual information between a subregion in one boundary and the whole of the other boundary.

The key finite temperature effect is that, for small regions, there is a competition between different possible minimal surfaces in the bulk homologous to $A$.  For $S(A)$, we need the smaller of $l(\gamma_A)$, the length of the minimal (connected) geodesic $\gamma_A$ homotopic to $A$, or  $l(\gamma_{A^c}) + L$,  where $\gamma_{A^c}$ is the minimal (connected) geodesic homotopic to $A^c$ and $L$ is the length of the closed geodesic at the horizon.   Similarly $S(A^c)$ is determined by either $l(\gamma_{A_c})$ or $l(\gamma_{A}) + L$, and there is an interesting competition between these two possibilities when $A$ is nearly the whole boundary.

\begin{figure}
\centering
\begin{subfigure}[t]{0.47\textwidth}
\centering
\includegraphics[scale=0.65]{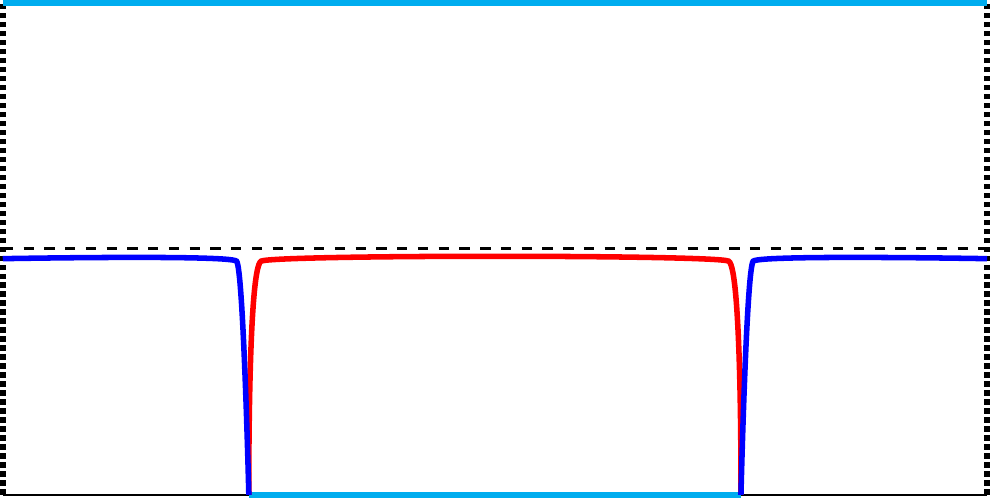}
\end{subfigure}
\begin{subfigure}[t]{0.47\textwidth}
\centering
\includegraphics[scale=0.65]{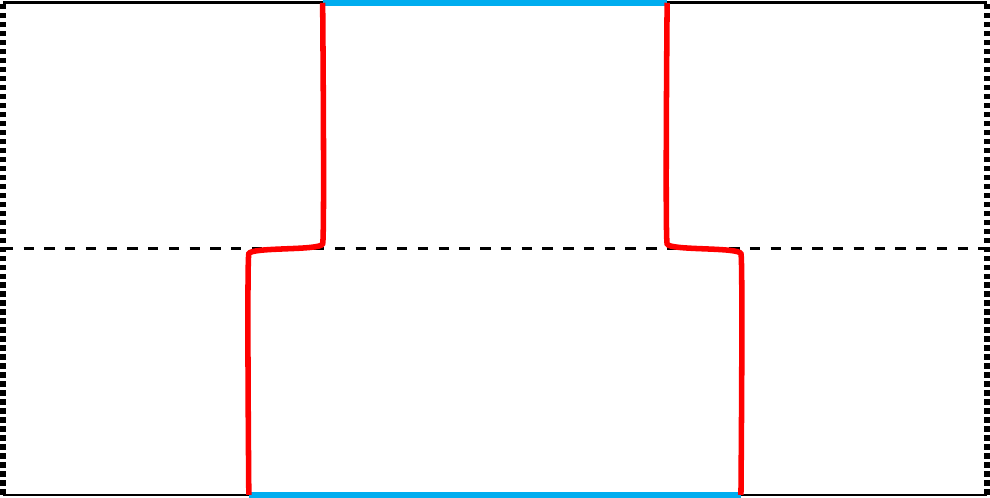}
\end{subfigure}
	\caption{The geodesics giving, in the high-temperature limit, the phases of entanglement entropy of the union of a pair of intervals (cyan)  lying on opposite boundaries, along with the event horizon added to satisfy the homology constraint, marked by the horizontal dashed line. When one of the intervals is a whole boundary, there are only two relevant phases (left), otherwise a third phase (right) may dominate, for which the corresponding geodesics cross the horizon and have endpoints lying on opposite boundaries. }\label{BTZphases}
\end{figure}

At high temperature the geodesics $\gamma_{A}, \gamma_{A_c}$ behave as shown in figure \ref{BTZphases}.  As a function of the angle $\phi$,  they drop quickly from the boundary to the horizon, hug the horizon while traversing an angle nearly $2\pi - \Delta \phi$, and then quickly return to the boundary.  One thus finds
\begin{equation}
\label{eq:lengths}
l(\gamma_{A}) = L \frac{\Delta \phi}{2 \pi} +O(L^0), \ \ \
l(\gamma_{A^c}) = L \left(1 - \frac{\Delta \phi}{2 \pi}\right) +O(L^0),
\end{equation}
which reproduces the behaviour in \eqref{Ilin} found in \cite{Morrison:2012iz}.

\begin{figure}
\centering
\begin{subfigure}[t]{0.47\textwidth}
\centering
\includegraphics[width=.8\textwidth]{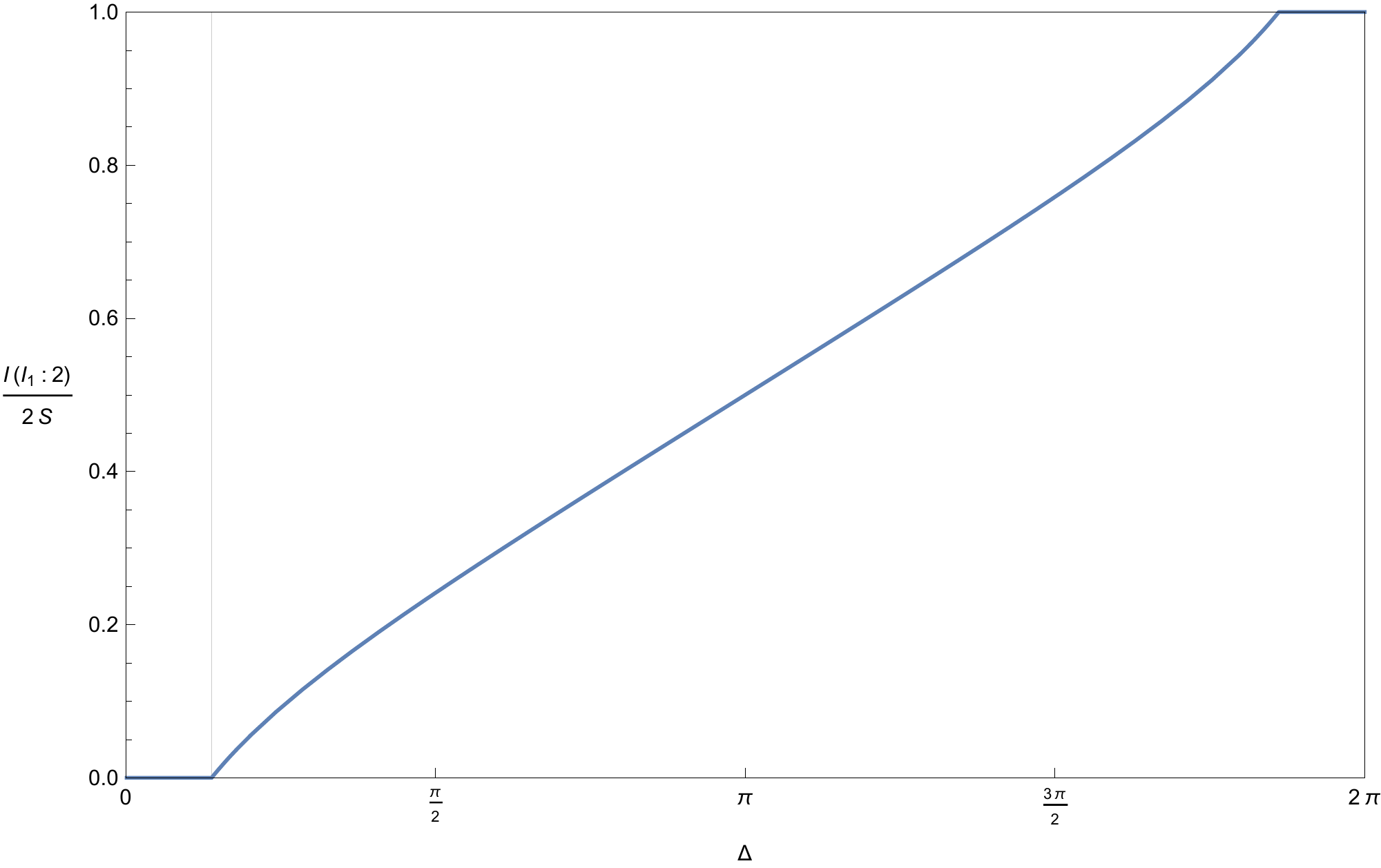}
\end{subfigure}
\begin{subfigure}[t]{0.47\textwidth}
\centering
\includegraphics[width=.8\textwidth]{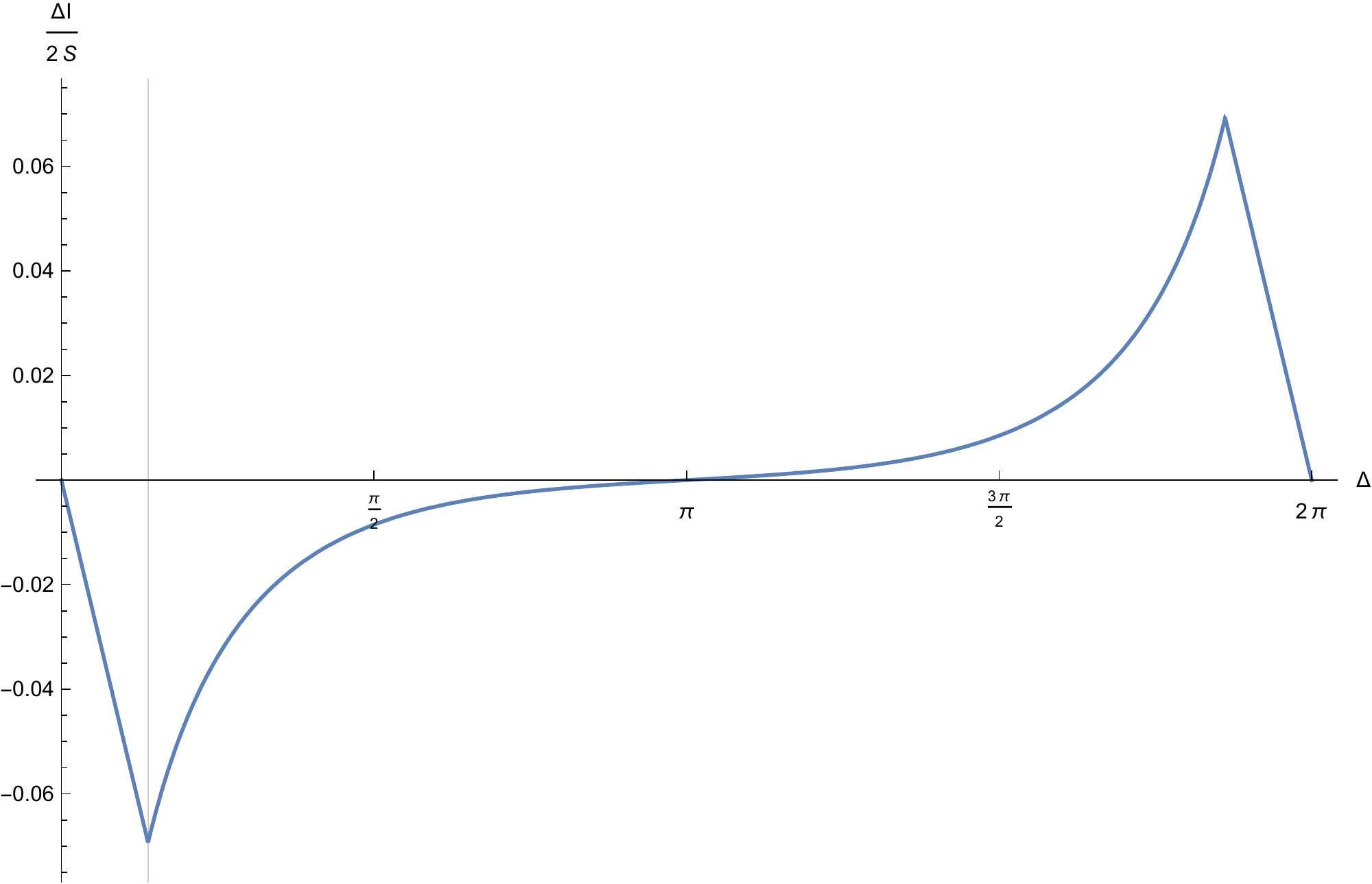}
\end{subfigure}
\caption{Left: Plot for $L=10$ of the mutual information $I(I_1;2)$ between an interval $I_1$ of size $\Delta$ in $B_1$ and all of $B_2$, as a function of $\Delta$, scaled by the maximal value $2S$. The mutual information increases approximately linearly in $\Delta$ and becomes non-zero at the phase transition where $\Delta \sim l^{-1}$ (vertical line near left edge). Right: Plot of the deviation in the mutual information shown at left from the high-temperature estimate $2S\frac{\Delta}{2\pi}$. This deviation is very small; for readability the vertical scale has been magnified relative to the left-hand plot.  The deviation is most significant for small and large values of $\Delta$, and decays exponentially in $L$ at intermediate values as expected.}
\label{fig:mi2}
\end{figure}

At finite $L$, corrections to \eqref{eq:lengths} are exponentially small in $L$ when $A$ and $A^c$ are larger than the thermal scale,  and the entropy remains close to linear in $\Delta \phi$.  But for any finite $L$ there is a Ryu-Takayanagi phase transition when either $A$ or $A^c$ becomes sufficiently small.  In that regime the relevant entropy $S(A)$ or $S(A^c)$ becomes controlled by the disconnected geodesic.  Thus, when the length of $A$ falls below $2\pi\log2/L + O(e^{-L}/L)$ one finds $I(A:B)= 0$.  For $A^c$ smaller than this threshold, one finds $I(A:B) = 2 S(B) = \frac{L}{2G}.$   Plotting the full $I(A:B)$ at large but finite $T$ clearly shows these ``plateaux" as in figure \ref{fig:mi2}. These plateaux were studied in \cite{Hubeny:2013gta}; they can be characterised in terms of saturation of the Araki-Lieb inequality as discussed in \cite{Headrick:2013zda}.

The fact that small intervals do not capture the entanglement with boundary 2 indicates that this information is encoded in a way that is non-local on the thermal length scale.   This is not surprising, but it is useful to note that this correction to the large $L$ picture has a natural expression in the language of  \cite{Almheiri:2014lwa}.  It says that at finite temperature the information about the entanglement between the two boundaries is not encoded locally in degrees of freedom at individual spatial points, but rather in a code subspace in each boundary, which entangles individual spatial degrees of freedom on the thermal scale. The ability to recover all of the information from any sufficiently large spatial subset of the degrees of freedom is the characteristic signature of such encoding in a code subspace. In \cite{Almheiri:2014lwa}, the size of the region needed to access information in a code subspace was related to the radial location of the bulk region encoded.  Similarly, in BTZ this size is related to the radial position of the horizon.

On a related note, the plateau at large $\Delta \phi$ appears precisely when the Ryu-Takayanagi surface for region $A$ is $\gamma(A^c)$ plus the horizon.  In other words, it occurs precisely when the so-called entanglement wedge \cite{Czech:2012be} -- the region inside this Ryu-Takayanagi surface -- reaches all the way to the horizon.  Indeed, in this case we see that it touches each and every point on the horizon and on $A$'s side of the horizon it misses only a small part of the space near $A^c$.  This suggests that the bulk near-horizon degrees of freedom are encoded non-locally in the CFT in such a way that they can be perfectly recovered from a large spatial subset $A$ that remains slightly smaller than the entire boundary.

\subsection{Pair of pants}

\begin{figure}
\centering
\begin{subfigure}[t]{0.2\textwidth}
\centering
\includegraphics[width=\textwidth]{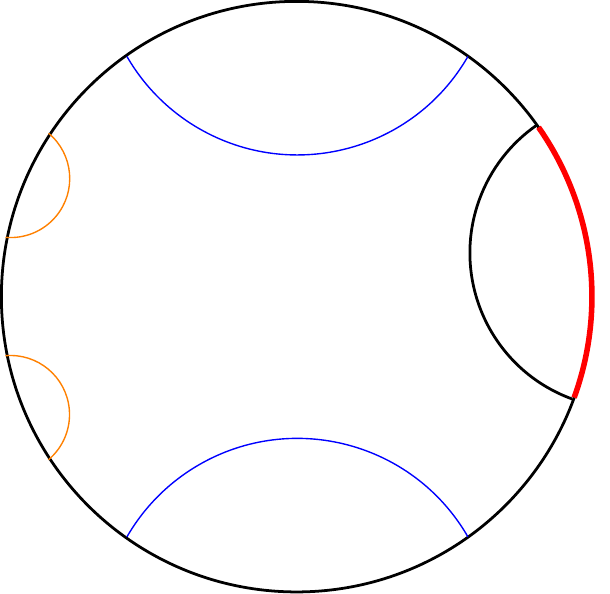}
\caption{Phase 1: $1$}
\end{subfigure}
\hspace{.07\textwidth}
\begin{subfigure}[t]{0.2\textwidth}
\centering
\includegraphics[width=\textwidth]{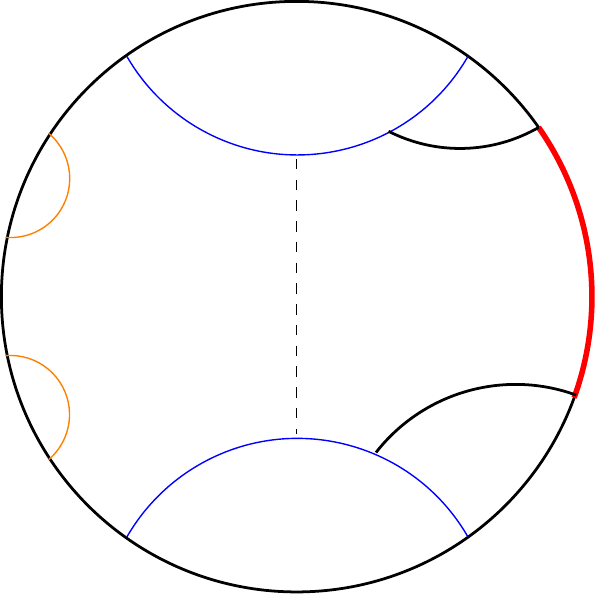}
\caption{Phase 2: $g_1^{-1}$}
\end{subfigure}\\
\begin{subfigure}[t]{0.2\textwidth}
\centering
\includegraphics[width=\textwidth]{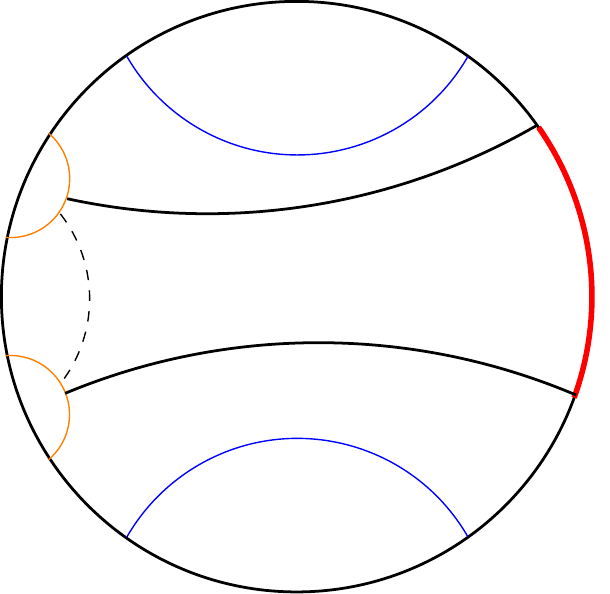}
\caption{Phase 3: $g_2^{-1}$}
\end{subfigure}
\hspace{.07\textwidth}
\begin{subfigure}[t]{0.2\textwidth}
\centering
\includegraphics[width=\textwidth]{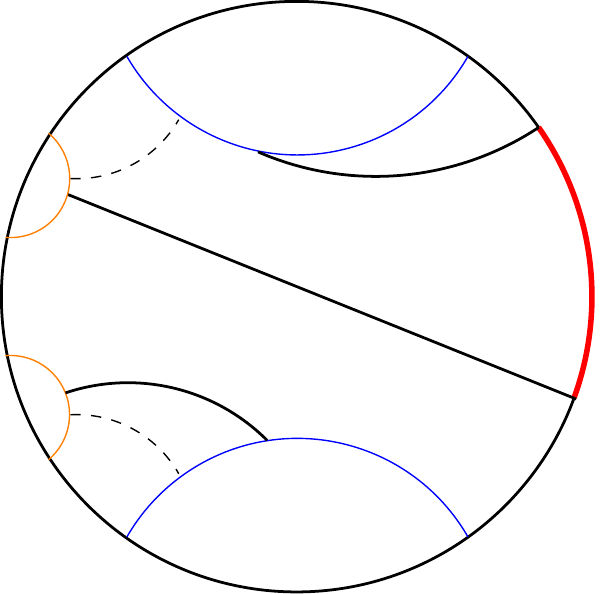}
\caption{Phase 4: $g_2g_1^{-1}$}
\end{subfigure}
\caption{The geodesics giving the four possible phases of entanglement entropy of a single interval, in red, along with the event horizons added to satisfy the homology constraint, marked by dashed lines.}\label{fig:3wormholePhases}
\end{figure}

\begin{figure}
\centering
\begin{subfigure}[t]{0.2\textwidth}
\centering
\includegraphics[width=\textwidth]{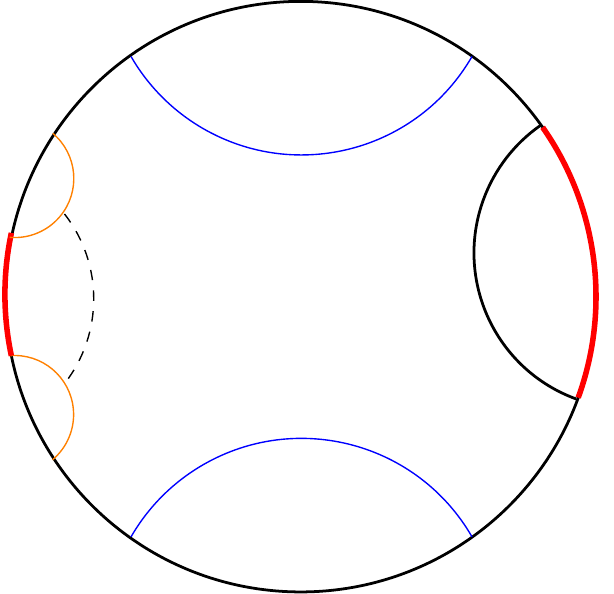}
\caption{Phase 1: $1$}
\end{subfigure}
\hspace{.07\textwidth}
\begin{subfigure}[t]{0.2\textwidth}
\centering
\includegraphics[width=\textwidth]{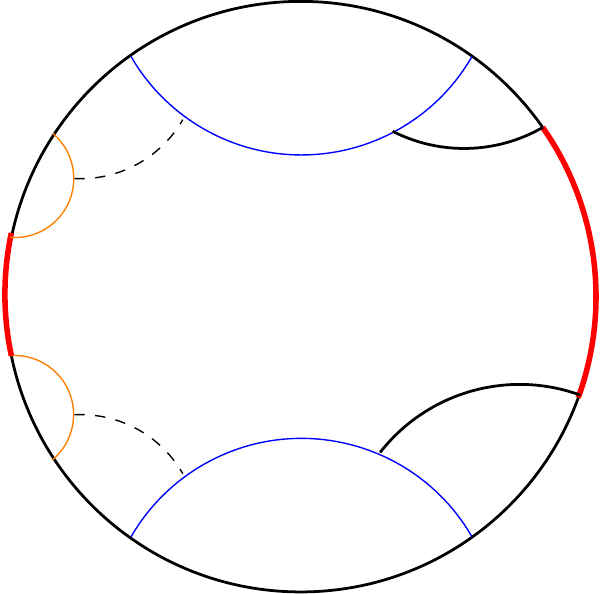}
\caption{Phase 2: $g_1^{-1}$}
\end{subfigure}\\
\begin{subfigure}[t]{0.2\textwidth}
\centering
\includegraphics[width=\textwidth]{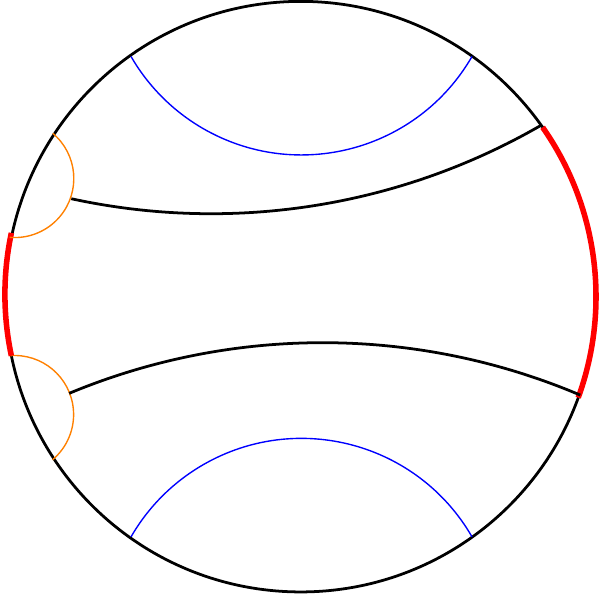}
\caption{Phase 3: $g_2^{-1}$}
\end{subfigure}
\hspace{.07\textwidth}
\begin{subfigure}[t]{0.2\textwidth}
\centering
\includegraphics[width=\textwidth]{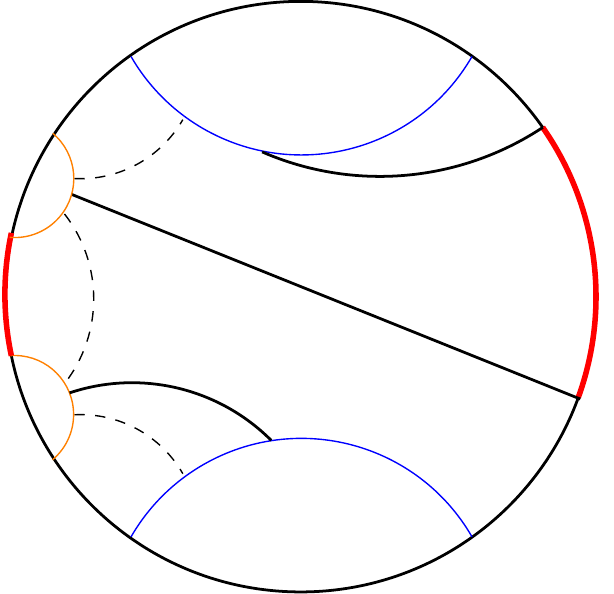}
\caption{Phase 4: $g_2g_1^{-1}$}
\end{subfigure}
\caption{The geodesics giving the four possible phases of entanglement entropy of the union of a single interval $A_1$ and the whole of boundary 2, in red, along with the event horizons added to satisfy the homology constraint, marked by dashed lines.}\label{fig:3wormholePhasesA1_2}
\end{figure}

For the pair of pants, we again study finite temperature corrections by considering the departure of the mutual information between a region in one boundary, say boundary 3, and the whole of another boundary, say boundary 1,
\be
I(A_3:1) = S_{A_3 }+ S_1 - S_{A_3 \cup 1},
\ee
from the approximation suggested by \eqref{eq:MIhighT}. As for the two boundary case, we expect the main departure to come near transitions between different phases, where different geodesics are exchanging dominance in the calculation of the holographic entanglement entropies. For this case, the phase transitions depend on two parameters: the size of $A_3$ and its location on boundary $3$. The different possible phases for $S(A_3)$ and $S(A_3 \cup B_1)$ are illustrated in figures \ref{fig:3wormholePhases} and \ref{fig:3wormholePhasesA1_2} respectively.

The calculation of the associated geodesic lengths can be easily automated using the description of the geodesic lengths as traces of corresponding $SL(2, \mathbb{R})$ group elements exploited in \cite{Maxfield:2014kra}. The lengths of the relevant geodesics can be found by computing the appropriate matrix products and traces. While the exact form of the answer is complicated and unilluminating, the general structure is fairly simple, being built mostly from polynomials in parameters encoding horizon lengths and the position of the interval. With a list of all contributing monomials in hand, finding the length in the large $L$ limit is equivalent to finding the maximum of a set of linear functions. This calculation is implemented in \texttt{Mathematica} by performing a truncated series expansion.

\begin{figure}
	\centering
	\includegraphics[width=.5\textwidth]{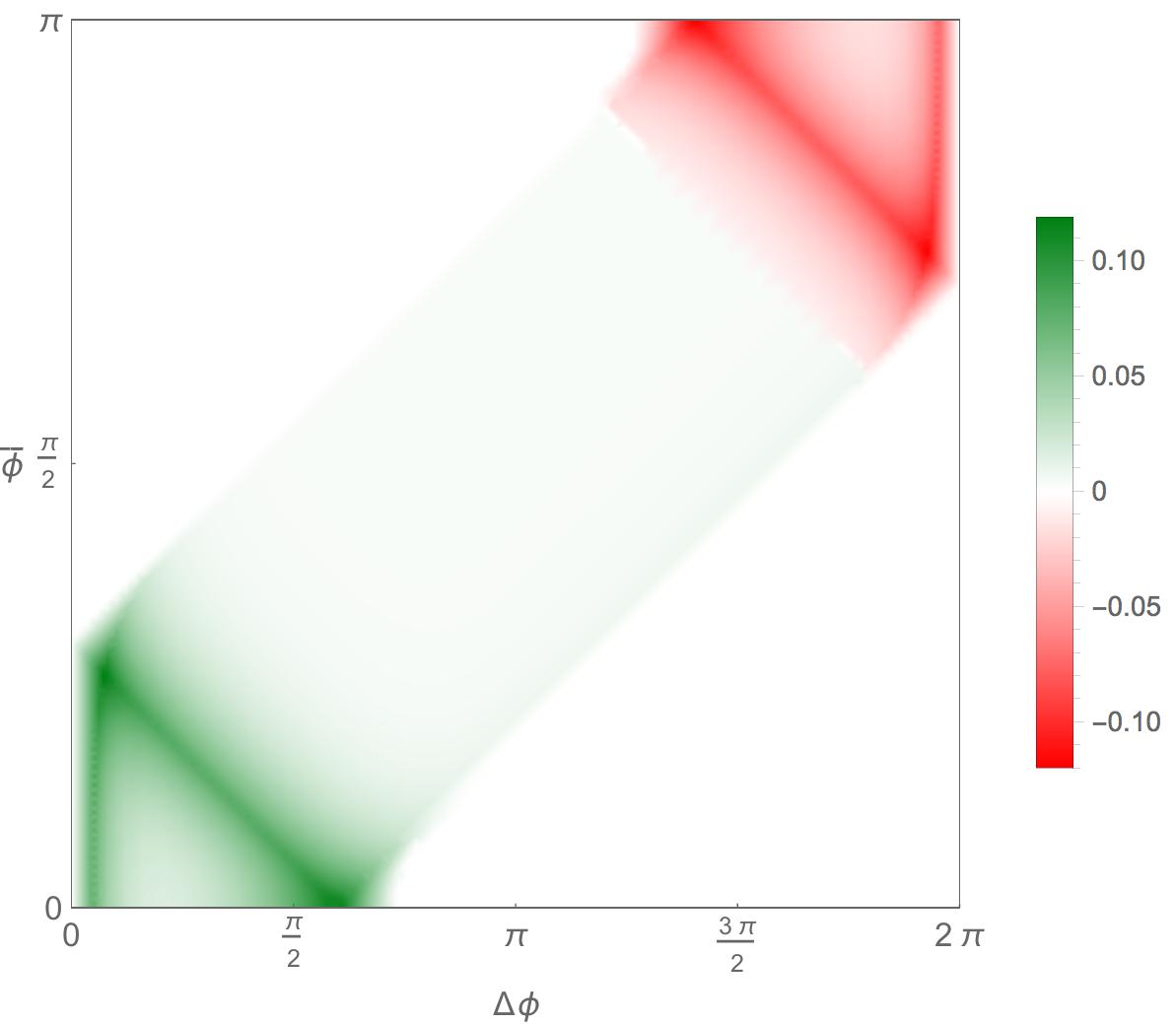}
	\caption{Deviation of the mutual information $I(A:B_1)$ between a region $A = \{\phi_3 \in [\bar{\phi} - \Delta \phi/2, \bar{\phi} + \Delta \phi/2] \}$ in $B_3$ and the whole of boundary $1$ from the piecewise-linear form implied by figure \ref{fig:sew} (lower right) for  $L_1=9,L_2=15,L_3=30$. 
	 We plot the ratio between the error and the maximal mutual information (twice the entropy of $B_1$). Here $\phi_3 = 0$ is the leftmost point in figure \ref{fig:sew} (lower right) and for comparison the inverse temperature $\beta_3$ is $\frac{(2\pi)^2}{30}\approx 1.3$.  The error is exponentially small in $L$, except in a region of thermal scale around certain phase transitions, where the order $L^0$ terms in \eqref{eq:MIhighT} contribute. The diagonal lines with largest error occur where an endpoint of $A$ leaves the region of $B_3$ entangled with $B_1$. The vertical lines with similarly large error are along a plateau phase transition, as occurs in the two boundary case.} \label{MIerrors}
\end{figure}

 Of course, the full series can also be computed numerically. The results are summarized in figure \ref{MIerrors}, which shows numerical results at finite-temperature for deviations from the high-temperature approximation \eqref{eq:MIhighT}.  The errors are indeed largest near the regions where nearby horizons are not exponentially close (i.e., where the causal shadows become large) and for intervals of size comparable to the thermal scale.  Such regimes are close to phase transitions in the mutual information, where pairs of minimal curves exchange dominance.

In addition to the bipartite entanglement, at large but finite temperature one expects to find tripartite entanglement associated with the shadow region between the horizons. But as noted above the area of the pair-of-pants causal shadow is $A_{CS} = 2\pi$ in AdS units for all values of the moduli $L_a$. Chopping off the exponentially thin ``arms,'' it can be useful to model this region as an AdS-scale disk.  This is quite reminiscent of the picture obtained in the tensor network model of the AdS vacuum in \cite{Pastawski:2015qua}, where different spatial regions mostly had bipartite entanglement, with a residual multipartite component corresponding to an AdS-scale region.

\section{Discussion}
\label{disc}

Our main result is that, in the limit of large black hole horizons, the pair-of-pants wormhole in 2+1 gravity is dual to a CFT state formed by sewing together thermofield doubles in one of the manners shown on the right of figure \ref{fig:sew}, or to the degenerate case that interpolates between them.  We showed this  by directly analyzing the CFT path integral defining the state $|\Sigma\rangle$, and used consistency with bulk holographic calculations of the mutual information to argue that the $\Sigma$-wormhole dominates the associated bulk Euclidean path integral.  We focused on the pair of pants for simplicity but -- as will be discussed further below -- it is easy to extend the central aspects of our  discussion to more complicated wormhole spacetimes.

We also focused on the case of circular boundaries, but the same conclusions apply to the planar case.  In 2+1 bulk dimensions, such high-temperature $n$-boundary planar cases are just AdS$_3$ in non-standard coordinates corresponding to performing certain conformal transformations on the dual CFT vacuum that are singular at $n$ points on the circle, with each segment running between two such singular points representing a distinct planar boundary.  One may also consider wormholes having both planar and circular boundaries.

Let us now briefly describe the extension to more general Riemann surfaces. Recall that a general orientable Riemann surface  $\Sigma$ (other than the sphere or annulus) can be decomposed into pairs-of-pants.  Let us think of $\Sigma$ as the $t=0$ slice of a wormhole spacetime with $n$ boundaries each asymptotic to AdS$_3$.  Then the surface contains geodesics $H_i$ ($i=1,2,\dots,n$) that define bifurcation surfaces of the event horizons for each boundary.  In addition, it contains a number of internal geodesics.  Each pair-of-pants decomposition of $\Sigma$ corresponds to cutting $\Sigma$ into pair-of-pants pieces along some set of these internal geodesics.  It will be convenient for us to also cut along the $H_i$ so that we in fact decompose $\Sigma$ into $n$ cylinders $C_i$ and some number of pair-of-pants pieces $\Sigma_I$. In a somewhat redundant notation, we will refer to the geodesics forming the three boundaries of $\Sigma_I$ as $H_{Ia}$ for $a=1,2,3$.  Note that the set of $H_{Ia}$ includes the horizons $H_i$. In this decomposition, the moduli space of the Riemann surface is parametrised by the lengths $L_{Ia}$ of the $H_{Ia}$ and the twists $\theta_{Ia}$ specifying the relative rotation between the two pairs of pants on the internal geodesics where we are sewing pairs of pants together.

\begin{figure}
\begin{subfigure}[t]{0.34\textwidth}
\centering
\includegraphics[scale=0.7]{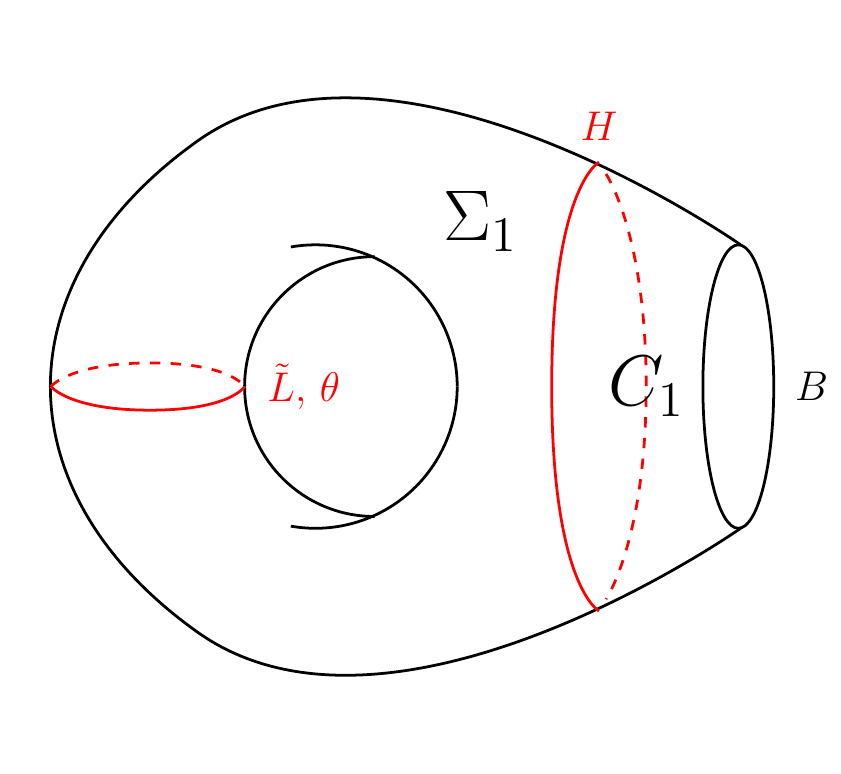}
\end{subfigure}
\begin{subfigure}[t]{0.25\textwidth}
\centering
\raisebox{20pt}{\includegraphics[scale=0.6]{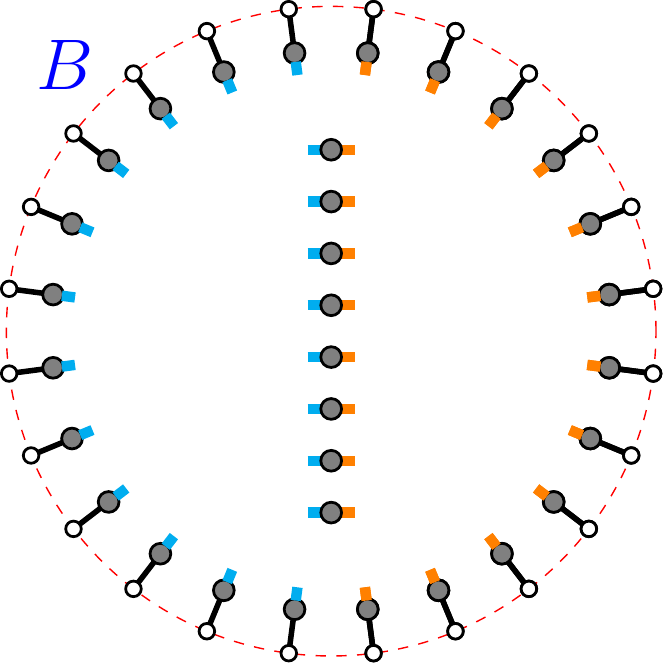}}
\end{subfigure}
\begin{subfigure}[t]{0.28\textwidth}
\centering
\raisebox{33pt}{\includegraphics[scale=0.6]{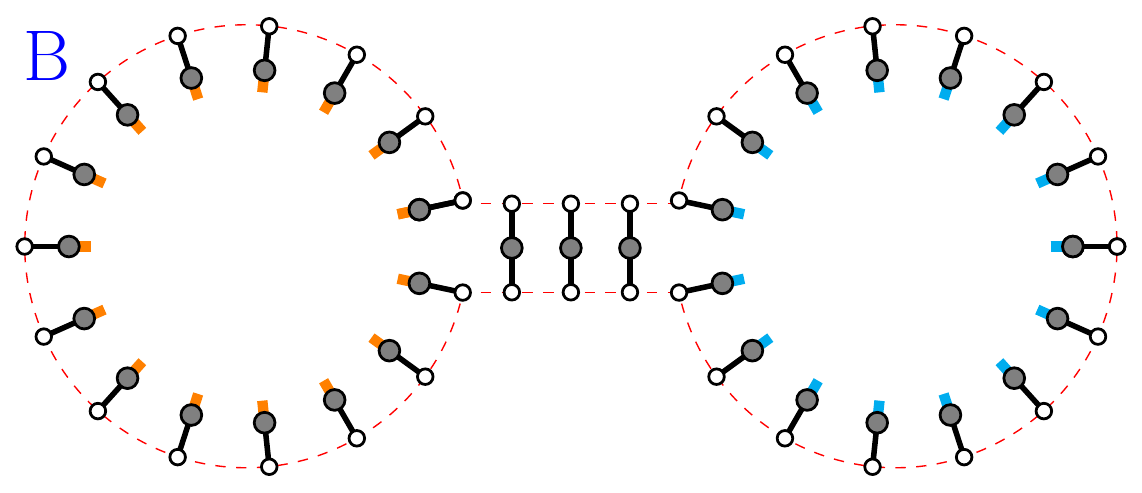}}
\end{subfigure}
\caption{A decomposition of the torus wormhole (left) with one boundary $B$  into a single pair of pants $\Sigma_1$ and a single cylinder $C_1$. Taking boundary 3 of $\Sigma_1$ to adjoin $C_1$, we see that there are three distinct moduli: the length $L_3$ of $H$, $L_1=L_2=\tilde{L}$, and a possible twist $\theta$. Tensor networks for $\theta = 0$ dual CFT states with large $L_3, \tilde L$ are also shown for $L_3 < 2 \tilde{L}$ (middle) and $L_3 > 2 \tilde{L}$ (right). In both cases, corresponding cyan and orange links are to be identified as dictated by the twist angle $\theta$.  For $\theta =0$, this identification is reflection about the vertical axis through the center of each diagram.  (Without this reflection, the spacetime is a punctured Klein bottle instead of a torus.)  The cyan and orange links should be viewed as exponentially short, while the black links have length $\beta/2$ set by the inverse effective temperature $\beta$ of the black hole.  So for $\theta =0$ these identifications generate exponentially short closed loops which  can be removed from the tensor without changing the state at leading order in large $L_3, \tilde L$ and central charge $c$. See discussion in main text below. }
\label{fig:decomposition_torus}
\end{figure}

\begin{figure}
\centering
\begin{subfigure}[t]{0.4\textwidth}
\centering
\includegraphics[width=0.7\textwidth]{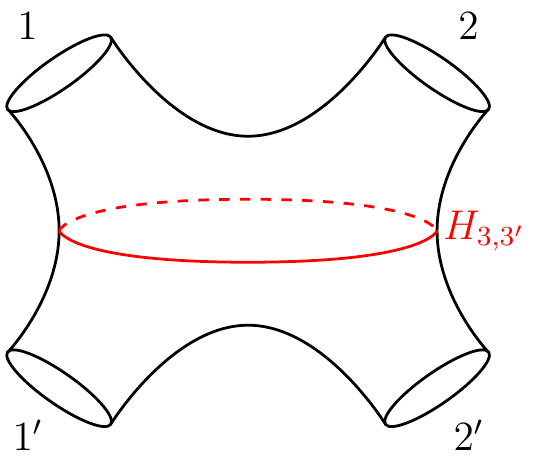}
\caption{$\theta = 0$}
\end{subfigure}
\begin{subfigure}[t]{0.4\textwidth}
\centering
\includegraphics[width=0.7\textwidth]{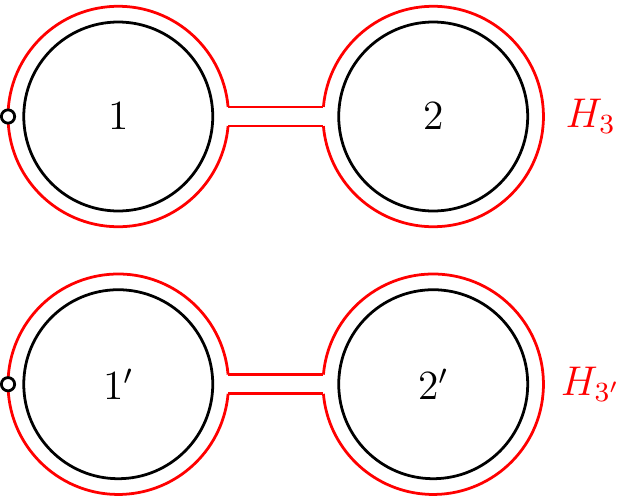}
\end{subfigure}
\begin{subfigure}[t]{0.4\textwidth}
\centering
\includegraphics[width=0.7\textwidth]{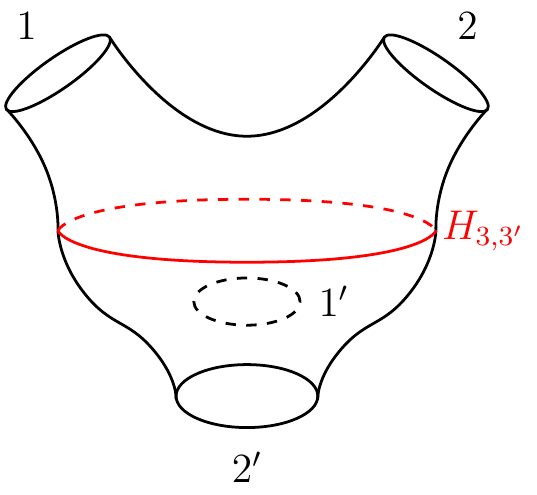}
\caption{$\theta = \frac{\pi}{2} $}
\end{subfigure}
\begin{subfigure}[t]{0.4\textwidth}
\centering
\includegraphics[width=0.7\textwidth]{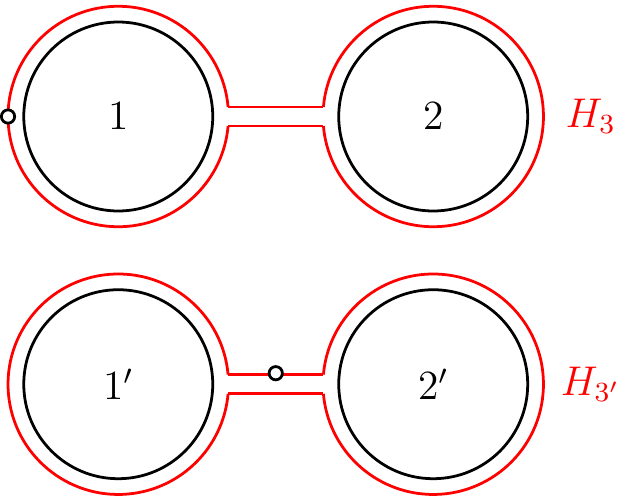}
\end{subfigure}
\begin{subfigure}[t]{0.4\textwidth}
\centering
\includegraphics[width=0.7\textwidth]{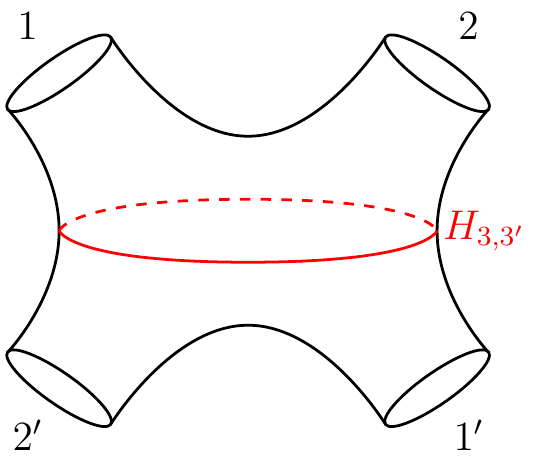}
\caption{$\theta = \pi$}
\end{subfigure}
\begin{subfigure}[t]{0.4\textwidth}
\centering
\includegraphics[width=0.7\textwidth]{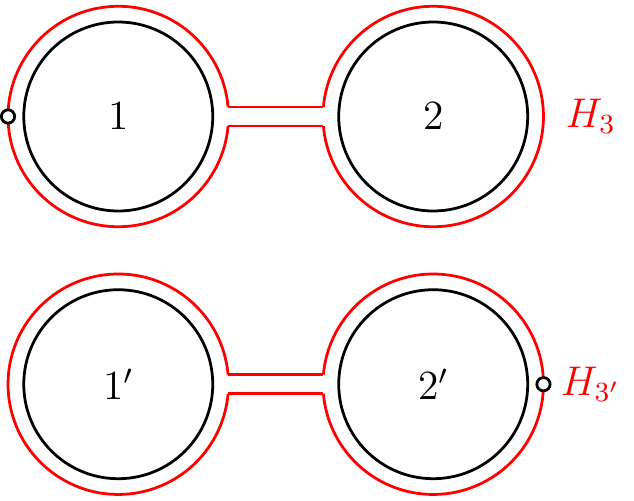}
\end{subfigure}
\caption{(Left diagrams) Two pairs-of-pants $\Sigma$ and $\Sigma'$ are each cut along $H_{3,3'}$ (red) and the pair of interior portions are sewn together along the cuts with twists $\theta=0,\pi/2,\pi$ to form a four-boundary wormhole. Here we consider the case with $L_3 > L_1+ L_2$, $L_{3'}>L_{1'}+L_{2'}$. In (a,c left) each pair of pants is bisected by an additional closed geodesic (not shown) that runs vertically around the diagram.  In the high-temperature limit, the corresponding entanglement structure is given by identifiying the outer boundaries $H_{3, 3'}$ of a pair of ``eyeglass" diagrams, shown in red  on the right-hand figure. This identification entails a twist $\theta$ which is represented by the dot in each cut which are identified across the join in accordance with the twist. For $\theta=0$ (top) we infer that boundaries $1$ and $2$ are each entangled only with $1'$ and $2'$ respectively. For $\theta=\frac{\pi}{2}$ (middle) and the chosen values of $L_a$ intervals within any given boundary are entangled with intervals in each of the others.   The pattern of such entanglements become chaotic at generic $\theta$, though a twist in this setting never entangles two distinct intervals within the same boundary. For $\theta=\pi$ (bottom) boundaries $1$ and $2$ are each entangled only with $2'$ and $1'$ respectively. }
\label{fig:4B}
\end{figure}

Each $\Sigma_I$ has the same geometry as the causal shadow region lying between the three horizons in figure \ref{fig:BTZrep} as defined by the corresponding $L_{Ia}$.  So each $\Sigma_\pm$ has area precisely $2\pi$, independent of moduli. Any $H_{Ia}$ which is long will lie exponentially close to another $H_{Ia}$ (or another region of the same $H_{Ia}$) across the causal shadow region. As a result, a large number of such $\Sigma_I$ can be sewn together without introducing an appreciable causal shadow or an appreciable reduction of the local energy density along each boundary.  Away from the special points in each boundary corresponding to vertices in our previous discussion, the effective $|TFD\rangle$ temperature remains uniform in the round conformal frame specified by the cylinders $C_i$.  Note that this is needed for consistency with the fact that the solution is precisely BTZ outside each horizon $H_i$, so that each boundary has constant energy density in our round frame\footnote{When the number of such pieces becomes comparable to the lengths $L_i$ of the horizons $H_i$, the qualitative effect on the CFT state $|\Sigma\rangle$ will depend on how these pieces of causal shadow are distributed along each boundary, and in particular on whether any regions of the boundaries do in fact remain far enough away to retain their $|TFD \rangle$ description.}. Some simple examples are shown in figures  \ref{fig:decomposition_torus} and \ref{fig:4B}, the former being a 1-boundary wormhole whose causal shadow at $t=0$ has the topology of a punctured torus.

Interesting new behaviour can arise as a function of the twists as we sew together pairs of pants with the structure of the lower panel in figure \ref{fig:sew}, as we illustrate by example in figure \ref{fig:4B}.   Consider for example a four-boundary wormhole with external horizons $H_1$, $H_2$, $H_1'$, $H_2'$, and split it into two pairs of pants along an internal geodesic $H_3= H_3'$ separating $H_{1,2}$ from $H_{1,2}'$.  We take $L_3 \gg L_1 + L_2$, $L_3 \gg L_1' + L_2'$. If $H_3$ and $H_3'$ are identified via some twist $\theta$, a given region in say $H_1$ is entangled with a region in $H_3$, which is in turn identified with some region in $H_3'$. For generic $\theta$ at large $L_3$ this will be entangled with some other region in $H_3'$, which is then identified back to $H_3$. For large $L_3$ we will cycle through the identification between $H_3$ and $H_3'$ many times before finally identifying the region with one of the other horizons ($H_2,H_{1'},H_{2'}$). In the limit where $L_3$ is much larger than the external horizons, the identifications resulting from a general twist are chaotic and appear to give a fractal entanglement structure.   It would be interesting to characterize these structures in more detail, and to relate this behavior to the well-known chaotic dynamics of geodesics on compact hyperbolic spaces.

\begin{figure}
\centering
\includegraphics[scale=0.8]{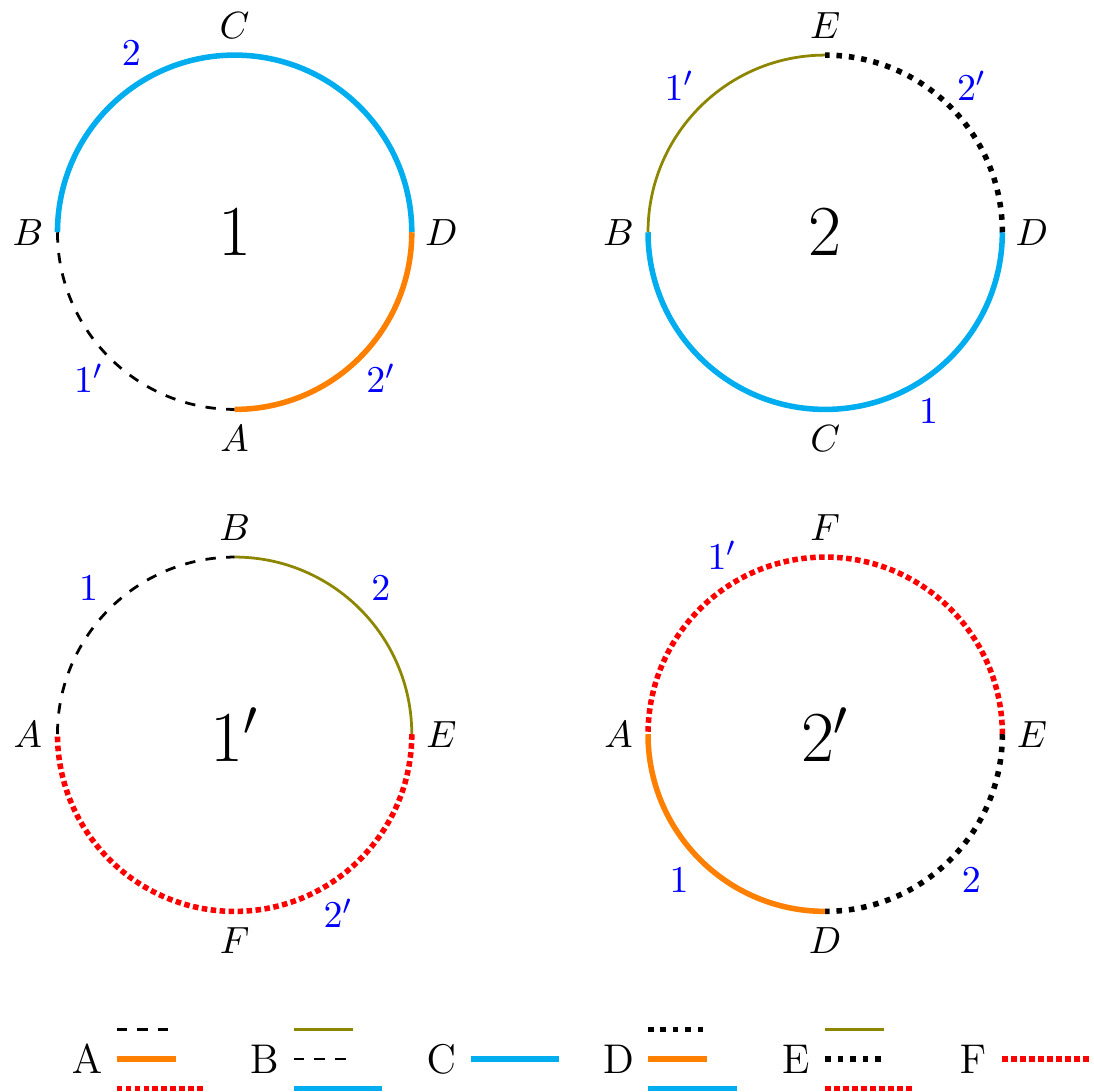}
\caption{The entanglement structure of the 4-boundary wormhole (figure \ref{fig:4B}) with large $L_3=L_{3'}$ for $L_1 = L_{1'} = L_2 = L_{2'} = \frac{1}{4} L_3$ and $\theta=\frac{11\pi}{8}$.
The state is well-described by a tensor network analogous to figure \ref{fig:sew} right.  The state on any pair of boundary intervals formatted in the same way (color, dots/dashes/solid lines) is a local piece of $|TFD\rangle$; the labels $1,1',2,2'$ indicate the boundaries connected to each TFD segment. The TFD intervals join at four vertices $A,B,D,E$ located as shown.  Each vertex connects the 3 local TFD states listed in the key below the diagram.  $C,F$ are not vertices, but are simply in the middle of the indicated (relatively long) TFD intervals. Some TFD strips connect oppositely-oriented intervals, while some preserve orientation.}
\label{fig:4Bexample}
\end{figure}

Another subtlety arises in cases like that shown in figure \ref{fig:4B} $a)$, where we consider the four-boundary system with zero twist, and take for simplicity $L_1 = L_1'$ and $L_2 = L_2'$ with $L_3$ again very large. Then sewing together the two copies of figure \ref{fig:sew} (lower right) indicates that $B_1$ is entangled only with $B_{1'}$ and that $B_2$ is entangled only with $B_{2'}$. As a result, taking $A = B_1 \cup B_{1'}$ and $B = B_2 \cup B_{2'}$, the CFT state has $I(A:B)=0$ (at leading order in large $c$ and $L$).  This result may seem is surprising from the bulk point of view, as $\Sigma$ contains a closed geodesic that runs vertically around figure \ref{fig:4B} (a), separating $A$ from $B$. So HRT predicts $I(A:B) = L/2G$, where $L$ is the length of this geodesic.  This would be consistent with the above prediction as large $L_3$ makes this geodesic exponentially short so that its length can be ignored at leading order.  Note that this geodesic is short only for zero twist: we saw that for small-but-nonzero twist $\theta$ a part of $B_1$ is instead entangled in a local $|TFD\rangle$ state with part of $B_{2'}$, so the mutual information is non-zero and grows as we scale up the $L_a$.   Thus $A$ and $B$ can no longer be separated by a closed bulk geodesic of negligible length\footnote{This is also clear from the fact that the local bulk geometry of these regions is essentially that of a segment of BTZ, and any such separating geodesic must traverse part of this segment of BTZ and thus have non-negligible length}.

Despite the above consistency, the appearance of such a short geodesic also suggests that our $\Sigma$-wormhole may not in fact be the dominant bulk saddle for the CFT state $|\Sigma \rangle$.  It seems natural to conjecture that -- unless forbidden by global features like a choice of spin structure --  when $\Sigma$ contains such exponentially short geodesics there will be another bulk saddle of smaller action where the geometry is modified so that these geodesics are contractible when viewed as living on the boundary of the new saddle. That is, we conjecture that $|\Sigma \rangle$ in such cases is in fact dual to a bulk geometry with $t=0$ surface $\Sigma'$ built by cutting $\Sigma$ along all exponentially small geodesics and capping off the resulting holes with small disks. This $\Sigma'-$ ``wormhole'' may not then connect all the boundaries.  From the tensor network point of view, the point is that the network obtained by gluing together two copies of figure \ref{fig:sew} with no twist breaks up into two disconnected components, one connecting $B_1$ and $B_{1'}$ and one connecting $B_2$ and $B_{2'}$.  The remaining chains merely form closed loops. At leading order in large central charge $c$ the properties of the state $|\Sigma \rangle$ are unchanged if we remove all chains that form closed loops rather than ending on boundaries.  The resulting tensor network defines the manifold $\Sigma'$.  The difference between
$|\Sigma \rangle$ and  $|\Sigma'\rangle$ is then exponentially small at large $c$, and we conjecture the $\Sigma'$-``wormhole'' to be the leading bulk saddle describing both states. This feature also arises for the punctured torus shown in figure \ref{fig:decomposition_torus} for $L_3 < 2 \tilde L$ (middle figure) with vanishing twist $\theta$.

It is worth elaborating further on this last point.  As noted in the caption for figure \ref{fig:sew}, the diagrams in this paper include only a simple cartoon of the $|TFD\rangle$ tensor networks from e.g. \cite{Maldacena:2013xja}.  The full tensor network for $|\Sigma \rangle$ obtained by sewing together $|TFD\rangle$ pieces as we describe will be correspondingly more complicated as well.  In particular, returning to the simple example of two pairs of pants with very large $L_3$ sewn together along the corresponding boundary, this full tensor network will certainly not factorize into unentangled states on $B_1B_{1'}$ and $B_2 B_{2'}$.   Instead, it will merely imply that the mutual information between $B_1B_{1'}$ and $B_2 B_{2'}$ remains of order $1$ at large central charge $c$.  This is analogous to $|TFD\rangle$ below the Hawking-page transition where it describes two entangled copies of a thermal gas on pure global AdS$_3$ backgrounds. Our conjecture is thus that the dominant bulk geometry at $t=0$ is correctly predicted by removing parts of the full tensor network that fail to transmit mutual information of order $c$.  We note that evaluating this criterion requires understanding the tensor structure of each node in the tensor network implied by the CFT dynamics; it is not apparent from the graph representation of the tensor network alone.

\begin{figure}
\centering
\includegraphics[scale=0.8]{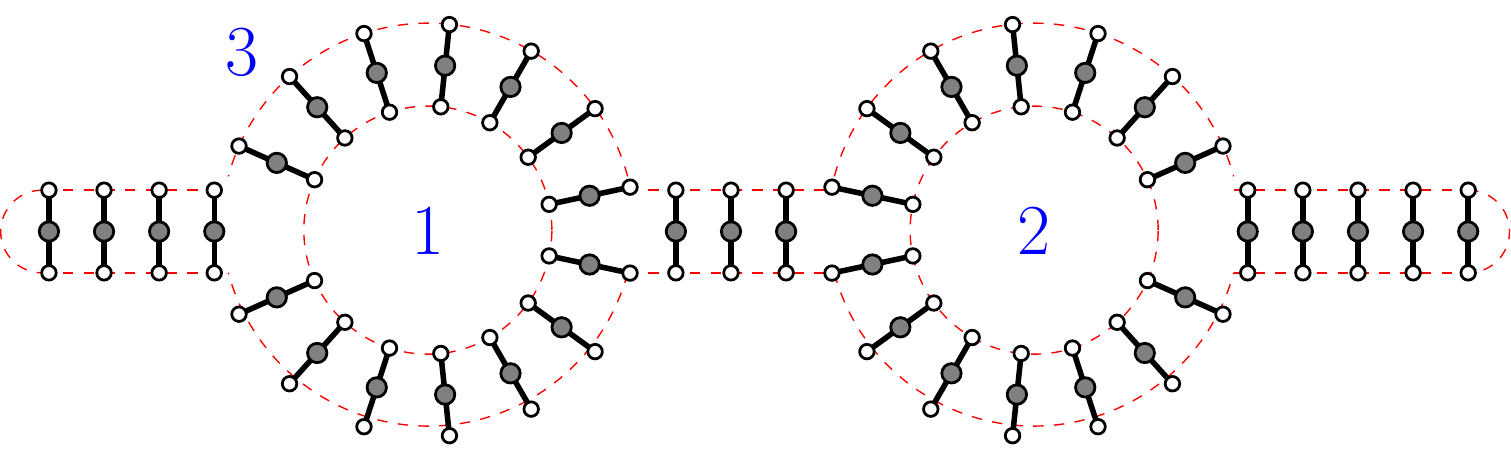}
\caption{Another way to sew three $|TFD\rangle$ states together.   Here the outer $|TFD\rangle$ has been sewn to itself at the ends as well as in the middle.  The sewing at the ends creates features we call `buds.'  Such buds are removed if one transforms the result to the round conformal frame.  One may construct similar buds from the vacuum by applying a smooth conformal transformation approximating over some region the singular one that gives the infinitely long planar thermofield double state.}
\label{fig:buds}
\end{figure}

So far we have considered tensor networks constructed by sewing together pair of pants networks in the way suggested by bulk wormhole geometries. But it is possible to consider a more general class of states defined by sewing together high-temperature $|TFD\rangle$ states in arbitrary fashions.  For example, one may sew a  $|TFD\rangle$ to itself (or others) so as to introduce a `bud'  on the tensor network as shown in figure \ref{fig:buds}.  Second, some pieces of some $|TFD\rangle$'s -- or even entire such states -- may now be entirely internal to the tensor network, lengthening some chains and thus lowering the local temperature.  In general, the chain length can then be non-uniform across any boundary.  Together, these two effects recover the freedom to make arbitrary conformal transformations relative to the round conformal frame used above. That is, these more general states must be related to the states considered above  rewritten in a more general conformal frame.

Finally, one may also generate non-orientable $\Sigma$ by performing antipodal identifications on some circle boundary.  For example, doing so one one boundary of a cylinder shows that the CFT state dual to the high-temperature AdS$_3$ geon (see e.g. \cite{Louko:1998hc}) is given by the thermofield double tensor network on a M\"obius strip.\footnote{This creates a local connection between antipodal points on the boundary. The fact that the bulk geodesic between antipodal points is short in the large temperature limit can be seen from the explicit formula for the geodesic lengths in \cite{Louko:2000tp}, although it is incorrectly stated there that the length of the geodesic through the identification is always longer than the one outside the horizon.}  The M\"obius strip can of course be constructed by cutting open the cylinder along $\phi =0$ and gluing the two ends back together with a half twist.  It is an interesting question to what extent such gluing operations reproduce desired states when applied to particular e.g. MERA-like tensor network representations of states at finite temperature and finite central charge $c$.  Results related to this issue will appear in \cite{Stanford}.

While we have stressed the limit where all $L_{a}$ become large, the discussion may be generalized to allow some $L_a$ to remain small.  The pair-of-pants CFT states $|\Sigma\rangle$ are then described by figure \ref{fig:sew} (right) with the small-$L$ boundaries contracted to points that merge with the vertices where the approximation by local TFDs breaks down.  But regions of any large-$L$ boundaries far from the new vertices remain well-described by the indicated local TFDs.  One should be aware that, due to the possibility of bulk phase transitions like those described above, having some $L_a$ small  may make it less clear which bulk spacetime is in fact dual to $|\Sigma\rangle$.  Nevertheless, the local TFD description of $|\Sigma \rangle$ remains valid.  In particular, any entanglement of large-$L$ boundaries with those having small-$L$ will be confined to intervals no longer than the effective thermal scale.  The tensor network issue dual to uncertainties regarding bulk connectivity is that some new vertices may now be trivial in the sense that they no longer lead to order-$c$ mutual information with the small boundary.  When this occurs and creates a `bud' as in figure \ref{fig:buds}, the bud may again be absorbed into a neighboring vertex without changing the large-$c$ structure of the state other than by acting with a conformal transformation\footnote{It is then the diagram without the bud that describes the round conformal frame.}.   Similar comments apply to $\Sigma$ having more boundaries or more general topology when some of the $L_{Ia}$ remain small.

Although we have discussed 2+1-dimensional bulk geometries above, but many of our considerations clearly apply to the higher dimensional case as well.  In particular, sewing together high-temperature $|TFD\rangle$'s defines a zoo of interesting states $|\Sigma\rangle$ and conformal geometries $\Sigma$.    And it is again natural to conjecture the CFT states $|\Sigma \rangle$ defined by such sewing operations to be dual to $\Sigma'$-wormholes defined by having a moment of time-reflection symmetry on which the induced geometry differs from (planar) Schwarzschild-AdS only by small corrections outside a finite number of AdS-scale regions.  But much remains to be understood and the details will prove interesting to explore.  In particular, one would like to find an algorithm that takes the tensor network naturally associated with $|\Sigma\rangle$ defined by the above gluing procedure and turns it into one in which the geometry of $\Sigma'$ is manifest -- e.g., with the tensor network providing a cellular decomposition of $\Sigma'$ in terms of AdS-scale cells \cite{Swingle:2009bg,2011JSP...145..891E,Swingle:2012wq,Hartman:2013qma}. One wonders if solving the Euclidean Einstein equations to construct $\Sigma'$ from $\Sigma$ can be related to a renormalization-group flow on tensor networks akin to those discussed in \cite{EvenblyVidal2014arXiv1412.0732E,EvenblyVidal2015arXiv150205385E}.

\hrulefill
\paragraph{Acknowledgements}
We thank Alex Maloney for helpful discussions.
D.M. was supported by the National Science Foundation under grant number PHY12-05500 and by funds
from the University of California. HM would like to thank UCSB for their hospitality during the project, and is supported by STFC, through a studentship and STEP award. AP is supported by an STFC studentship. SFR was supported by STFC under grant number ST/L000407/1. D.M. and S.F.R. thank the Aspen Center and its NSF Grant \#1066293 for their hospitality during the discussions where certain aspects of this project were conceived.  D. M.  also thanks the KITP for their hospitality during the final stages of the project, where his work was also supported in part by National Science foundation grant number PHY11-25915.

\appendix
\section{The horizons $H_1, H_2$ in BTZ coordinates}
\label{app:H1H2}

We now compute the parameters characterising the region $\Sigma_+$, corresponding to half of the $t=0$ slice of the three-boundary wormhole with horizons $H_1,H_2,H_3$, of respective lengths $L_1,L_2,L_3$. In the main text we ordered the lengths so that $L_3>L_1,L_2$; this assumption is relaxed here. The region $\Sigma_+$ is bounded by three geodesics $G_{ab}$, running between the boundary components labelled by $a$ and $b$, and meeting horizons $H_a,H_b$ orthogonally.

We use the metric \eqref{eq:BTZ} with $H_3$ lying at $\rho=0$ and $G_{13},G_{23}$ lying at $x=\pm \frac{L_3}{4}$.  Thus $x\in[-\frac{L_3}{4},\frac{L_3}{4}]$.  Consider a geodesic parameterised by arclength $s$, using a dot to denote differentiation with respect to $s$. From translation invariance, there is a conserved quantity $(1+\rho^2)\dot{x}$, which for geodesics with both endpoints at $\rho=\infty$ is given by $\sqrt{1+\rho_0^2}$, where $\rho_0>0$ is the minimal value of $\rho$. The geodesic is then given by

\begin{equation}
\rho = \rho_0 \cosh s,\quad x = x_0 + \tanh^{-1}\left(\frac{\tanh s}{\sqrt{1+\rho_0^2}}\right).
\end{equation}

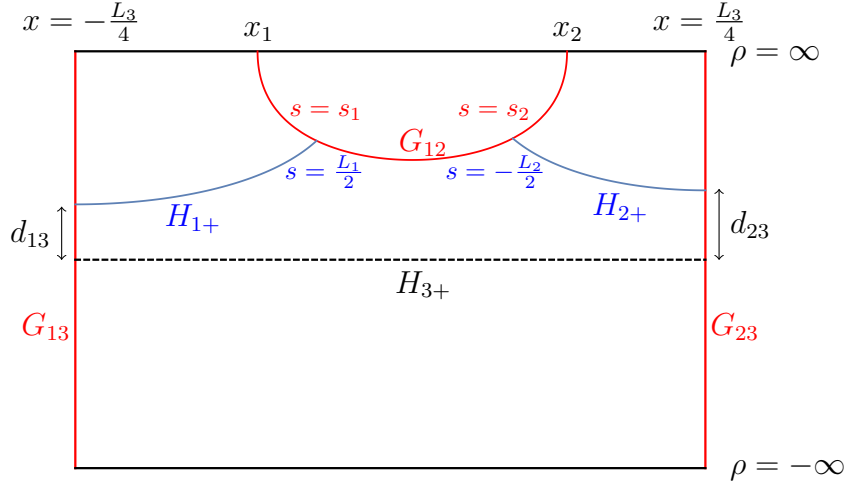
\begin{figure}
\centering
	\begin{tikzpicture}
    \node[anchor=south west,inner sep=0] (image) at (0,0,0) {\includegraphics[width=.5\textwidth]{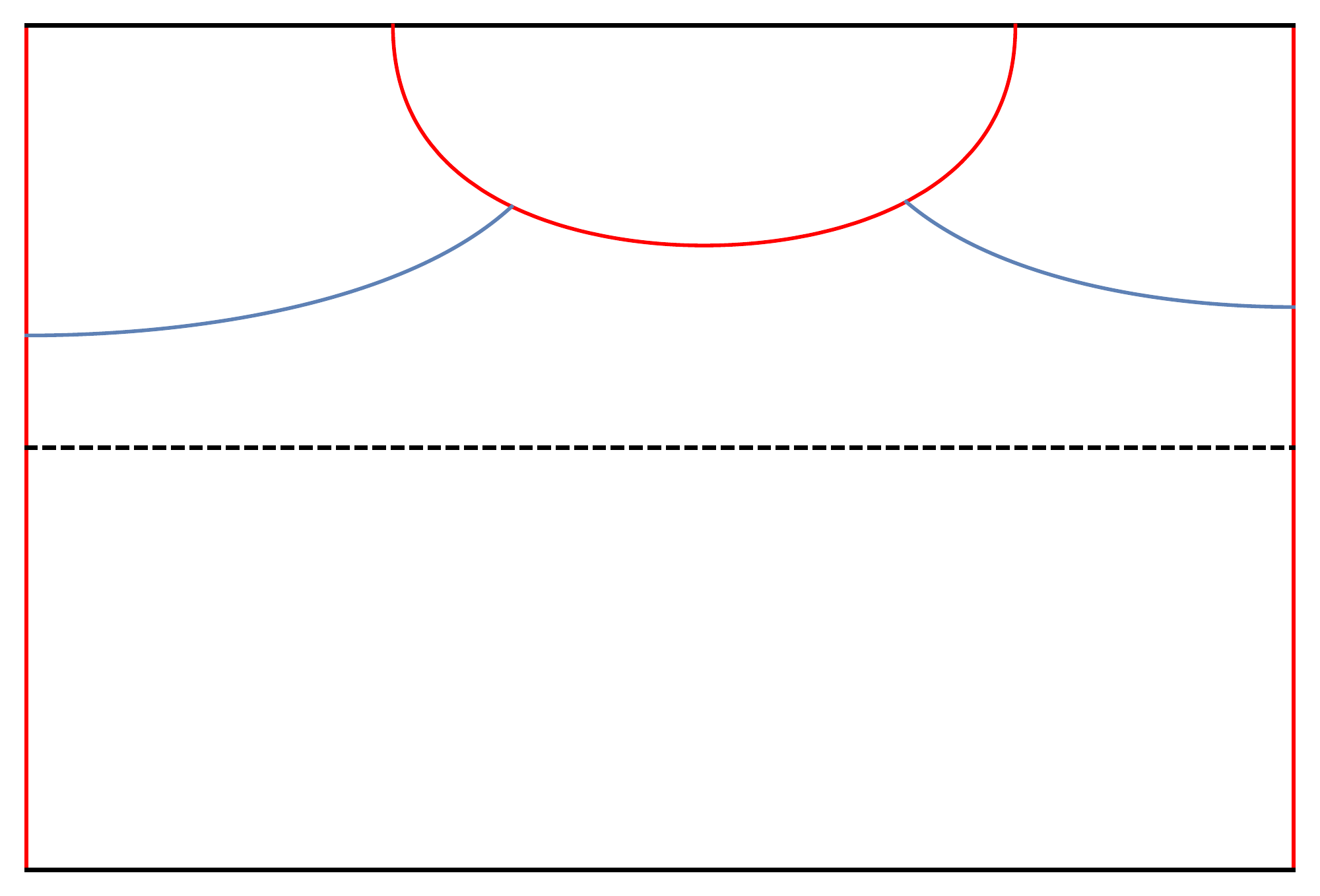}} ;
    \begin{scope}[x={(image.south east)},y={(image.north west)}]
        \node [below, blue] at (.2,.65) {$H_{1+}$};
        \node [below, blue] at (.85,.675) {$H_{2+}$};
        \node [below] at (.55,.5) {$H_{3+}$};
        \node [above, red] at (.55,.71) {$G_{12}$};
        \node [left, red] at (0.03,.35) {$G_{13}$};
        \node [right, red] at (.97,.35) {$G_{23}$};
        \node [above] at (.3,.97) {$x_1$};
        \node [above] at (.77,.97) {$x_2$};
        \draw [<->] (0,.5) -- (0,.62);
        \node [left] at (0,.56) {$d_{13}$};
        \draw [<->] (1,.5) -- (1,.66);
        \node [right] at (1,.58) {$d_{23}$};
        \node [right] at (1,.03) {$\rho = -\infty$};
        \node [right] at (1,.96) {$\rho = \infty$};
        \node [above] at (0.03,.97) {$x=-\frac{L_3}{4}$};
        \node [above] at (0.97,.97) {$x=\frac{L_3}{4}$};
        \node [above right, red] at (0.33,0.79) {\footnotesize$s=s_1$};
        \node [below, blue] at (0.40,0.76) {\footnotesize$s=\frac{L_1}{2}$};
        \node [above left, red] at (0.73,0.79) {\footnotesize$s=s_2$};
        \node [below, blue] at (0.66,0.76) {\footnotesize$s=-\frac{L_2}{2}$};
    \end{scope}
\end{tikzpicture}
\caption{The geodesics $G_{ab}$ bounding the patch $\Sigma_+$ in BTZ coordinates. The horizons $H_{a+}$ are also shown. The $\rho$ direction runs vertically, and $x$ horizontally. The positions where the geodesics intersect are labelled with the arclength along each curve, measured from the deepest point (minimal $\rho$), and $d_{13},d_{23}$ mark the minimal distances between the horizons.}\label{BTZstrip}
\end{figure}

Consider first the geodesics $G_{12}$ and that corresponding to $H_{1+}$  (see \cref{BTZstrip} for the various relevant geodesics and quantities). The endpoints of $G_{12}$ lie at $x=x_1,x_2$, and it will be convenient to parametrise these by the centre $\bar{x}=\frac{x_1+x_2}{2}$ and the half-width $\Delta x=\frac{x_2-x_1}{2}$. We intend to find $L_1$ in terms of these parameters, and along the way will also obtain the minimal distance $d_{13}$ between $H_{1+}$ and $H_{3+}$, as well as the position at which $G_{12}$ and $H_{1+}$ intersect.

The geodesics are given by
\begin{align}
	G_{12}: \rho &= \frac{\cosh s}{\sinh \Delta x},
	\quad x = \bar{x} + \tanh^{-1}\left(\tanh\Delta x\tanh s\right),
	\quad s\in\mathbb{R}, \\
	H_{1+}: \rho &= \sinh d_{13} \cosh s,
	\quad x = -\frac{L_3}{4} + \tanh^{-1}\left(\frac{\tanh s}{\cosh d_{13}}\right),
	\quad 0\leq s \leq \frac{L_1}{2},
\end{align}
with the constraint that they intersect at right angles at the endpoint of $H_{1+}$, where the arclength along $H_{1+}$ is $s=\frac{L_1}{2}$, and along $G_{12}$ is $s=s_1$, say, where $s_1<0$.

The condition that two geodesics intersect orthogonally determines the value of $\rho$ at which they meet in terms of the conserved quantities for the two geodesics; for $H_{1+}$ and $G_{12}$ it gives
\begin{equation}
	\rho^2 = \sinh^2 d_{13}+\coth^2\Delta x = \cosh^2 d_{13}+\csch^2\Delta x \text{  at intersection.}
\end{equation}
We now get two equations from identifying the value of $\rho$ at intersection with the values of $\rho$ for $G_{12}$ at $s=s_1$, and for $H_{1+}$ at $s=\frac{L_1}{2}$. A third comes from identifying the $x$ coordinates at these same arclengths:
\begin{align}
	\coth\Delta x &= \sinh d_{13} \sinh\frac{L_1}{2}, \\
	\sinh{s_1} &= -\sinh \Delta x \cosh d_{13}, \\
	\bar{x}+\tanh^{-1}(\tanh \Delta x \tanh s_1) &= -\frac{L_3}{4} + \tanh^{-1}\left(\frac{\tanh\frac{L_1}{2}}{\cosh d_{13}}\right).
\end{align}
We then solve for $L_1,d_{13},s_1$ in terms of $\bar{x},\Delta x$ to obtain
\begin{equation}
	\cosh\frac{L_1}{2} = \frac{\sinh\left(\frac{L_3}{4}+\bar{x}\right)}{\sinh \Delta x},\;
	\tanh d_{13} = \frac{\cosh \Delta x}{\cosh\left(\frac{L_3}{4}+\bar{x}\right)},\;
	\tanh s_1 = -\frac{\tanh \Delta x}{\tanh\left(\frac{L_3}{4}+\bar{x}\right)}.
\end{equation}
It is straightforward to translate these results into expressions for $L_2$, the distance $d_{23}$ between horizons $H_{2+}$ and $H_{3+}$, and $s_2$, the arclength along $G_{12}$ at which it intersects $H_{2+}$:
\begin{equation}
	\cosh\frac{L_2}{2} = \frac{\sinh\left(\frac{L_3}{4}-\bar{x}\right)}{\sinh \Delta x},\;
	\tanh d_{23} = \frac{\cosh \Delta x}{\cosh\left(\frac{L_3}{4}-\bar{x}\right)},\;
	\tanh s_2 = \frac{\tanh \Delta x}{\tanh\left(\frac{L_3}{4}-\bar{x}\right)}
\end{equation}

Finally, the above can be inverted to find $\bar{x}$, $\Delta x$, and $d_{ab}$ (where, in particular, $d_{12}=s_2-s_1$ is the minimal distance between $H_{1+},H_{2+}$) in terms of $L_1,L_2,L_3$.
\begin{align}
	\sinh\bar{x} &= \frac{\left(\cosh\frac{L_1}{2}-\cosh\frac{L_2}{2}\right)\sinh\frac{L_3}{4}}{\sqrt{\cosh^2\frac{L_1}{2}+\cosh^2\frac{L_2}{2}+2\cosh\frac{L_1}{2}\cosh\frac{L_2}{2}\cosh\frac{L_3}{2}}} , \\
	\sinh\Delta x &= \frac{\sinh\frac{L_3}{2}}{\sqrt{\cosh^2\frac{L_1}{2}+\cosh^2\frac{L_2}{2}+2\cosh\frac{L_1}{2}\cosh\frac{L_2}{2}\cosh\frac{L_3}{2}}} , \\
	\cosh d_{12} &= \frac{\cosh\frac{L_1}{2}\cosh\frac{L_2}{2}+\cosh\frac{L_3}{2}}{\sinh\frac{L_1}{2}\sinh\frac{L_2}{2}}\quad \text{(and permutations)}
\end{align}
In particular, the explicit inversion shows that the mapping between $(x_1,x_2)$ and $(L_1,L_2)$ is bijective and smooth.

We may now work out the asymptotic values of these quantities in the limit where all lengths $L_a$ are large. The typical expressions reduce to sums of exponentials of linear combinations of $L_a$, so there are separate regimes depending on the relative sizes of the exponents; these turn out to be three regimes where one horizon is longer than the sum of the others ($L_1>L_2+L_3$ and permutations), and the regime where no horizon is dominant in this way.

\begin{align}
	\bar{x} &\sim \begin{cases}
		\frac{L_3}{4}-\frac{1}{2}\exp\left(-\frac{L_1-L_2-L_3}{2}\right) & L_1>L_2+L_3 \\
		-\frac{L_3}{4}+\frac{1}{2}\exp\left(-\frac{L_2-L_1-L_3}{2}\right) & L_2>L_1+L_3 \\
		\frac{L_1-L_2}{4} & \text{otherwise}
	\end{cases}\\
	\Delta x &\sim \begin{cases}
		\exp\left(-\frac{L_1-L_3}{2}\right) & L_1>L_2+L_3 \\
		\exp\left(-\frac{L_2-L_3}{2}\right) & L_2>L_1+L_3 \\
		\frac{L_3-L_1-L_2}{4}+\log 2 & L_3>L_1+L_2 \\
		\exp\left(-\frac{L_1+L_2-L_3}{4}\right) & \text{otherwise}
	\end{cases}\\
	d_{12} &\sim \begin{cases}
		\frac{L_3-L_1-L_2}{2}+2\log2 & L_3>L_1+L_2 \\
		2 \exp\left(-\frac{L_2}{2}\right) & L_1>L_2+L_3 \\
		2 \exp\left(-\frac{L_1}{2}\right) & L_2>L_1+L_3 \\
		2 \exp\left(-\frac{L_1+L_2-L_3}{2}\right) & \text{otherwise}
	\end{cases}
\end{align}
The corrections in each case are exponentially small in the $L_a$, except when $L_3-L_1-L_2$ is order one, for example.

The interval $[x_1,x_2]$ looks qualitatively different in each of the four regimes. When $L_3>L_1+L_2$, it is long (the same order as the horizon lengths), and at a generic position. When $L_1>L_2+L_3$, it is exponentially short, and also close to the right end of the strip; it is similarly short and close to the left end when $L_2>L_1+L_3$. In the remaining regime, it is again exponentially short, but in a generic position.

\bibliographystyle{JHEP}
\bibliography{spatial}

\providecommand{\href}[2]{#2}\begingroup\raggedright\begin{thebibliography}{10}

\bibitem{Maldacena:2001kr}
J.~M. Maldacena, {\it {Eternal black holes in anti-de Sitter}},  {\em JHEP}
  {\bf 0304} (2003) 021, [\href{http://xxx.lanl.gov/abs/hep-th/0106112}{{\tt
  hep-th/0106112}}].

\bibitem{VanRaamsdonk:2010pw}
M.~Van~Raamsdonk, {\it Building up spacetime with quantum entanglement},  {\em
  General Relativity and Gravitation} {\bf 42} (2010), no.~10 2323--2329,
  [\href{http://xxx.lanl.gov/abs/1005.3035}{{\tt arXiv:1005.3035}}].

\bibitem{Czech:2012be}
B.~Czech, J.~L. Karczmarek, F.~Nogueira, and M.~Van~Raamsdonk, {\it Rindler
  quantum gravity},  {\em Classical and Quantum Gravity} {\bf 29} (2012),
  no.~23 235025, [\href{http://xxx.lanl.gov/abs/1206.1323}{{\tt
  arXiv:1206.1323}}].

\bibitem{Maldacena:2013xja}
J.~Maldacena and L.~Susskind, {\it Cool horizons for entangled black holes},
  {\em Fortschritte der Physik} {\bf 61} (2013), no.~9 781--811,
  [\href{http://xxx.lanl.gov/abs/1306.0533}{{\tt arXiv:1306.0533}}].

\bibitem{Hubeny:2007xt}
V.~E. Hubeny, M.~Rangamani, and T.~Takayanagi, {\it A covariant holographic
  entanglement entropy proposal},  {\em Journal of High Energy Physics} {\bf
  2007} (2007), no.~07 062,
  [\href{http://xxx.lanl.gov/abs/hep-th/0705.0016}{{\tt hep-th/0705.0016}}].

\bibitem{Lewkowycz:2013nqa}
A.~Lewkowycz and J.~Maldacena, {\it Generalized gravitational entropy},  {\em
  Journal of High Energy Physics} {\bf 2013} (2013), no.~8 1--29,
  [\href{http://xxx.lanl.gov/abs/1304.4926}{{\tt arXiv:1304.4926}}].

\bibitem{Ryu:2006bv}
S.~Ryu and T.~Takayanagi, {\it {Holographic derivation of entanglement entropy
  from AdS/CFT}},  {\em Phys.Rev.Lett.} {\bf 96} (2006) 181602,
  [\href{http://xxx.lanl.gov/abs/hep-th/0603001}{{\tt hep-th/0603001}}].

\bibitem{Marolf:2012xe}
D.~Marolf and A.~C. Wall, {\it {Eternal Black Holes and Superselection in
  AdS/CFT}},  {\em Class.Quant.Grav.} {\bf 30} (2013) 025001,
  [\href{http://xxx.lanl.gov/abs/1210.3590}{{\tt arXiv:1210.3590}}].

\bibitem{Morrison:2012iz}
I.~A. Morrison and M.~M. Roberts, {\it {Mutual information between thermo-field
  doubles and disconnected holographic boundaries}},  {\em JHEP} {\bf 1307}
  (2013) 081, [\href{http://xxx.lanl.gov/abs/1211.2887}{{\tt
  arXiv:1211.2887}}].

\bibitem{2009arXiv0912.1651V}
G.~{Vidal}, {\it {Entanglement Renormalization: an introduction}},  {\em ArXiv
  e-prints} (Dec., 2009) [\href{http://xxx.lanl.gov/abs/0912.1651}{{\tt
  arXiv:0912.1651}}].

\bibitem{Susskind:2014ira}
L.~Susskind, {\it {Computational Complexity and Black Hole Horizons}},
  \href{http://xxx.lanl.gov/abs/1403.5695}{{\tt arXiv:1403.5695}}.

\bibitem{Balasubramanian:2014hda}
V.~Balasubramanian, P.~Hayden, A.~Maloney, D.~Marolf, and S.~F. Ross, {\it
  {Multiboundary Wormholes and Holographic Entanglement}},  {\em
  Class.Quant.Grav.} {\bf 31} (2014) 185015,
  [\href{http://xxx.lanl.gov/abs/1406.2663}{{\tt arXiv:1406.2663}}].

\bibitem{Susskind:2014yaa}
L.~Susskind, {\it {ER=EPR, GHZ, and the Consistency of Quantum Measurements}},
  \href{http://xxx.lanl.gov/abs/1412.8483}{{\tt arXiv:1412.8483}}.

\bibitem{Hayden:2011ag}
P.~Hayden, M.~Headrick, and A.~Maloney, {\it {Holographic Mutual Information is
  Monogamous}},  {\em Phys.Rev.} {\bf D87} (2013), no.~4 046003,
  [\href{http://xxx.lanl.gov/abs/1107.2940}{{\tt arXiv:1107.2940}}].

\bibitem{Brill:1995jv}
D.~R. Brill, {\it {Multi - black hole geometries in (2+1)-dimensional
  gravity}},  {\em Phys.Rev.} {\bf D53} (1996) 4133--4176,
  [\href{http://xxx.lanl.gov/abs/gr-qc/9511022}{{\tt gr-qc/9511022}}].

\bibitem{Aminneborg:1997pz}
S.~Aminneborg, I.~Bengtsson, D.~Brill, S.~Holst, and P.~Peldan, {\it {Black
  holes and wormholes in (2+1)-dimensions}},  {\em Class.Quant.Grav.} {\bf 15}
  (1998) 627--644, [\href{http://xxx.lanl.gov/abs/gr-qc/9707036}{{\tt
  gr-qc/9707036}}].

\bibitem{Brill:1998pr}
D.~Brill, {\it {Black holes and wormholes in (2+1)-dimensions}},  {\em
  Lect.Notes Phys.} {\bf 537} (2000) 143,
  [\href{http://xxx.lanl.gov/abs/gr-qc/9904083}{{\tt gr-qc/9904083}}].

\bibitem{Aminneborg:1998si}
S.~Aminneborg, I.~Bengtsson, and S.~Holst, {\it {A Spinning anti-de Sitter
  wormhole}},  {\em Class.Quant.Grav.} {\bf 16} (1999) 363--382,
  [\href{http://xxx.lanl.gov/abs/gr-qc/9805028}{{\tt gr-qc/9805028}}].

\bibitem{Krasnov:2000zq}
K.~Krasnov, {\it {Holography and Riemann surfaces}},  {\em
  Adv.Theor.Math.Phys.} {\bf 4} (2000) 929--979,
  [\href{http://xxx.lanl.gov/abs/hep-th/0005106}{{\tt hep-th/0005106}}].

\bibitem{Krasnov:2003ye}
K.~Krasnov, {\it {Black hole thermodynamics and Riemann surfaces}},  {\em
  Class.Quant.Grav.} {\bf 20} (2003) 2235--2250,
  [\href{http://xxx.lanl.gov/abs/gr-qc/0302073}{{\tt gr-qc/0302073}}].

\bibitem{Skenderis:2009ju}
K.~Skenderis and B.~C. van Rees, {\it {Holography and wormholes in 2+1
  dimensions}},  {\em Commun.Math.Phys.} {\bf 301} (2011) 583--626,
  [\href{http://xxx.lanl.gov/abs/0912.2090}{{\tt arXiv:0912.2090}}].

\bibitem{Bao:2015bfa}
N.~Bao, S.~Nezami, H.~Ooguri, B.~Stoica, J.~Sully, et~al., {\it {The
  Holographic Entropy Cone}},  \href{http://xxx.lanl.gov/abs/1505.0783}{{\tt
  arXiv:1505.0783}}.

\bibitem{Pastawski:2015qua}
F.~Pastawski, B.~Yoshida, D.~Harlow, and J.~Preskill, {\it {Holographic quantum
  error-correcting codes: Toy models for the bulk/boundary correspondence}},
  \href{http://xxx.lanl.gov/abs/1503.0623}{{\tt arXiv:1503.0623}}.

\bibitem{Banados:1992gq}
M.~Banados, M.~Henneaux, C.~Teitelboim, and J.~Zanelli, {\it {Geometry of the
  (2+1) black hole}},  {\em Phys.Rev.} {\bf D48} (1993), no.~6 1506--1525,
  [\href{http://xxx.lanl.gov/abs/gr-qc/9302012}{{\tt gr-qc/9302012}}].

\bibitem{Friedman:1993ty}
J.~L. Friedman, K.~Schleich, and D.~M. Witt, {\it {Topological censorship}},
  {\em Phys.Rev.Lett.} {\bf 71} (1993) 1486--1489,
  [\href{http://xxx.lanl.gov/abs/gr-qc/9305017}{{\tt gr-qc/9305017}}].

\bibitem{Galloway:1999bp}
G.~Galloway, K.~Schleich, D.~Witt, and E.~Woolgar, {\it {Topological censorship
  and higher genus black holes}},  {\em Phys.Rev.} {\bf D60} (1999) 104039,
  [\href{http://xxx.lanl.gov/abs/gr-qc/9902061}{{\tt gr-qc/9902061}}].

\bibitem{Maxfield:2014kra}
H.~Maxfield, {\it {Entanglement entropy in three dimensional gravity}},
  \href{http://xxx.lanl.gov/abs/1412.0687}{{\tt arXiv:1412.0687}}.

\bibitem{Hubeny:2013gta}
V.~E. Hubeny, H.~Maxfield, M.~Rangamani, and E.~Tonni, {\it {Holographic
  entanglement plateaux}},  {\em JHEP} {\bf 1308} (2013) 092,
  [\href{http://xxx.lanl.gov/abs/1306.4004}{{\tt arXiv:1306.4004}}].

\bibitem{Headrick:2013zda}
M.~Headrick, {\it {General properties of holographic entanglement entropy}},
  {\em JHEP} {\bf 1403} (2014) 085,
  [\href{http://xxx.lanl.gov/abs/1312.6717}{{\tt arXiv:1312.6717}}].

\bibitem{Almheiri:2014lwa}
A.~Almheiri, X.~Dong, and D.~Harlow, {\it {Bulk Locality and Quantum Error
  Correction in AdS/CFT}},  \href{http://xxx.lanl.gov/abs/1411.7041}{{\tt
  arXiv:1411.7041}}.

\bibitem{Louko:1998hc}
J.~Louko and D.~Marolf, {\it {Single exterior black holes and the AdS / CFT
  conjecture}},  {\em Phys.Rev.} {\bf D59} (1999) 066002,
  [\href{http://xxx.lanl.gov/abs/hep-th/9808081}{{\tt hep-th/9808081}}].

\bibitem{Louko:2000tp}
J.~Louko, D.~Marolf, and S.~F. Ross, {\it {On geodesic propagators and black
  hole holography}},  {\em Phys.Rev.} {\bf D62} (2000) 044041,
  [\href{http://xxx.lanl.gov/abs/hep-th/0002111}{{\tt hep-th/0002111}}].

\bibitem{Stanford}
B.~Czech, G.~Evenbly, L.~Lamprou, S.~McCandlish, X.~Qi, J.~Sully, and G.~Vidal
  To appear.

\bibitem{Swingle:2009bg}
B.~Swingle, {\it {Entanglement Renormalization and Holography}},  {\em
  Phys.Rev.} {\bf D86} (2012) 065007,
  [\href{http://xxx.lanl.gov/abs/0905.1317}{{\tt arXiv:0905.1317}}].

\bibitem{2011JSP...145..891E}
G.~{Evenbly} and G.~{Vidal}, {\it {Tensor Network States and Geometry}},  {\em
  Journal of Statistical Physics} {\bf 145} (Nov., 2011) 891--918,
  [\href{http://xxx.lanl.gov/abs/1106.1082}{{\tt arXiv:1106.1082}}].

\bibitem{Swingle:2012wq}
B.~Swingle, {\it {Constructing holographic spacetimes using entanglement
  renormalization}},  \href{http://xxx.lanl.gov/abs/1209.3304}{{\tt
  arXiv:1209.3304}}.

\bibitem{Hartman:2013qma}
T.~Hartman and J.~Maldacena, {\it {Time Evolution of Entanglement Entropy from
  Black Hole Interiors}},  {\em JHEP} {\bf 1305} (2013) 014,
  [\href{http://xxx.lanl.gov/abs/1303.1080}{{\tt arXiv:1303.1080}}].

\bibitem{EvenblyVidal2014arXiv1412.0732E}
G.~{Evenbly} and G.~{Vidal}, {\it {Tensor Network Renormalization}},  {\em
  ArXiv e-prints} (Dec., 2014) [\href{http://xxx.lanl.gov/abs/1412.0732}{{\tt
  arXiv:1412.0732}}].

\bibitem{EvenblyVidal2015arXiv150205385E}
G.~{Evenbly} and G.~{Vidal}, {\it {Tensor network renormalization yields the
  multi-scale entanglement renormalization ansatz}},  {\em ArXiv e-prints}
  (Feb., 2015) [\href{http://xxx.lanl.gov/abs/1502.0538}{{\tt
  arXiv:1502.0538}}].

\end{thebibliography}\endgroup

\end{document}